\documentclass[%
 preprint,
 superscriptaddress,
 amsmath,amssymb,
aip,
citeautoscript,
]{revtex4-2}

\usepackage{physics}
\usepackage[dvipsnames]{xcolor}
\usepackage{hyperref}
\usepackage[section]{placeins}
\usepackage{graphicx}
\usepackage{dcolumn}
\usepackage{bm}

\usepackage{float}
\usepackage{booktabs}
\usepackage{multirow}

\begin{document}

\title{Discovery of sustainable energy materials via the machine-learned material space}

\author{Malte Grunert}
\email{malte.grunert@tu-ilmenau.de}
\affiliation{Institute of Physics and Institute of Micro- and Nanotechnologies, Technische Universit\"at Ilmenau, 98693 Ilmenau, Germany}
\affiliation{These authors contributed equally to this work.}
\author{Max Großmann}
\affiliation{Institute of Physics and Institute of Micro- and Nanotechnologies, Technische Universit\"at Ilmenau, 98693 Ilmenau, Germany}
\affiliation{These authors contributed equally to this work.}
\author{Erich Runge}
\affiliation{Institute of Physics and Institute of Micro- and Nanotechnologies, Technische Universit\"at Ilmenau, 98693 Ilmenau, Germany}

\date{\today}

\begin{abstract}
Does a machine learning model actually gain an understanding of the material space?
We answer this question in the affirmative on the example of the \textsc{OptiMate} model, a graph attention network trained to predict the optical properties of semiconductors and insulators.
By applying the UMAP dimensionality reduction technique to its latent embeddings, we demonstrate that the model captures a nuanced and interpretable representation of the materials space, reflecting chemical and physical principles, without any user-induced bias. 
This enables clustering of almost 10,000 materials based on optical properties and chemical similarities.
Beyond this understanding, we demonstrate how the learned material space can be used to identify more sustainable alternatives to critical materials in energy-related technologies, such as photovoltaics.
These findings demonstrate the dual utility of machine learning models in materials science: Accurately predicting material properties while providing insights into the underlying materials space.
The approach demonstrates the broader potential of leveraging learned materials spaces for the discovery and design of materials for diverse applications, and is easily applicable to any state-of-the-art machine learning model.
\end{abstract}

\maketitle

\section{Introduction}
Machine learning (ML) has undoubtedly taken the world of materials science by storm in recent years and has shown impressive results. 
These comprise the acceleration of existing theoretical applications (most notably molecular dynamics via machine-learned force fields \cite{Batatia2022Design, Batatia2022mace, MatterSim2024} and deep learning of Hamiltonians \cite{Li2022, Gong2023, Tang2024}) as well as the direct prediction of functional properties such as the band gap, \cite{Pilania2016, Lee2016, Xie2018, Rajan2018, Zhuo2018, Chen2019, Choudhary2021, Omee2022, Wang2022} the static dielectric constant, \cite{Morita2020, Takahashi2020} or the transition temperature of conventional superconductors. \cite{Stanev2018, Cerqueira2023, Sanna2024}
As another example, we \cite{Grunert2024b} and others \cite{Ibrahim2024, Hung2024} have recently presented models that predict frequency-dependent optical properties directly from the crystal structure in a matter of milliseconds. 
Such predictions are crucial for many energy-related applications such as photovoltaic systems, epsilon-near-zero materials, optical sensors, or energy-efficient light-emitting devices.
An open question is whether the models actually "learn" complex materials science concepts, as a trained and experienced materials scientist, chemist, or physicist would, or whether they approximate the mapping from structure to property, similar to what a density functional theory (DFT) code does.
If the first alternative were the case, then an additional question arises: Can this understanding of materials science in one domain be exploited in another?

In this work, we begin with answering the first question by investigating in detail the learned representations of a graph attention network (GAT) trained on a large materials database to predict the optical spectra of materials.
The considered model \textsc{OptiMate} \cite{Grunert2024b} consists, like most models used nowadays in materials science, of several parts of quite different structure, namely (i) an atom embedding multilayer perceptron (MLP) which in essence learns an adapted periodic system, (ii) a varying number of message passing blocks allowing for communication between atoms, and (iii) a spectra prediction MLP which converts the learned representation of the material into the optical spectra. 
In this work, we extract the high-dimensional latent embeddings, i.e., the internal representation for each material, at different points in the model.
At each level, we construct low-dimensional human-interpretable projections of them using the Unified Manifold Approximation Projection (UMAP) technique (we note that in this article we use the term UMAP both when referring to the algorithm itself and the resulting two-dimensional "map").\cite{UMAP2018}
UMAP is a nonlinear dimension-reduction technique that preserves local distances in the high-dimensional space while also aiming to preserve global distances as much as possible.\cite{UMAP2018}
This means that points that are close together in the high-dimensional space, and are thus treated similarly by the model, are placed close together in the low-dimensional space (usually two or three dimensions for visualization).
Briefly, the UMAP algorithm calculates similarities between points in the high-dimensional embeddings based on nearest-neighbor distances, calculates similarities in the lower dimension for a given transformation, and optimizes the transformation by minimizing the cross-entropy between the high-dimensional and the low-dimensional similarities.
UMAP has been used extensively (outside the ML community) in the medicine and biology community, where it is used, for example, to visualize various high-dimensional quantities often found in cell, genome, or protein studies.\cite{Becht2018, Karczewski2020, Sikkema2023, Lin2023}
We note in passing that using t-distributed stochastic neighbor embeddings (t-SNE) instead of UMAP to visualize the high-dimensional data yields similar results, see Supporting Information (SI).

The main result of these investigations is that the model learns internal representations that are remarkably similar to what an experienced materials scientist might create, but for over 10,000 materials across a large number of different material classes.
In essence, the model appears to learn the material space in terms of optical properties.

Apart from this perhaps surprising, but ultimately only "pleasing" result, we then proceed to answer the second question by showing how this representation of the material space can be put to practical use.
As a use case, we chose the important task of identifying alternative, more sustainable and less critical optical materials for energy-related technologies such as photovoltaics or solar hydrogen generation.
While these applications are critical for combating climate change, they often rely on materials which are not ideal due to various environmental, economic, or social concerns.
We use quantifications of various dimensions of these concerns for all main-group elements, namely the supply risk \cite{EuCrit} (termed "criticality" in accordance with the terminology of the European Union) and "sustainability",\cite{euchemsElementScarcity} by which is meant the amount of available reserves in relation to the current usage.
We emphasize that many competing concepts have been published that characterize materials quantitatively, e.g., as more or less 'critical', 'strategic', 'rare', 'sustainable', and so on.
Their analysis and comparison is beyond the scope of this work and must be done from an interdisciplinary perspective. 
The freely available \textsc{OptiMate} code and the code used to generate the UMAPs can easily be adapted to any quantifiable concept of 'problematic elements', for concepts and methods, see, e.g., Ref.~\onlinecite{Nuss2014} and \onlinecite{Schrijvers2020}.    
By using such criteria to color-code a UMAP of the material space, it is possible to identify possible alternative materials with similar optical properties and chemistry for these applications.

We analyze a particular example - III-V semiconductors used in solar cells - in detail in the results section and provide the UMAPs themselves in the SI and online as an interactive website.
There, compounds are displayed in a color that is a quantification of the criterion in question.
Anyone interested in a particular compound can easily explore the material space and find compounds in the neighborhood that \textsc{OptiMate} considers similar. 

\section{Results}
The model analyzed and exploited in this work is the \textsc{OptiMate} model \cite{Grunert2024b} developed and trained to predict the frequency-dependent dielectric function $\varepsilon(\omega)$.
Since we are not interested in benchmarking \textsc{OptiMate} in this work, but only in evaluating the nature and capabilities of its internal encodings, we train the model on the entire database presented in the original work, i.e., we do not perform a training-validation-test split of the dataset.
This does not significantly change the overall structure of the UMAPs obtained from the internal embeddings, but prevents a small number of materials in the test set from being misplaced in the constructed material space.
The model is sketched in Fig.~\ref{fig:UMAP_elems}(top left).
We extract the internal embeddings at various points in the model, namely, after the GATPool operation,\cite{Grunert2024b} termed $R_p$,  and after each intermediate step of the spectra prediction MLP, termed $R_i^{(1-3)}$.
In addition, we also extract the predicted spectra and treat it as an embedding, termed $R_s$, to contrast the information contained in it with the information contained in the embeddings of the earlier stages of the model.
As mentioned earlier, all of these embeddings are extremely high-dimensional (between 384- and 2001-dimensional) and as such are not humanly interpretable or even visualizable.
Therefore, we resort to UMAPs of these embeddings.
The UMAP parameters used in this work are described in the Methods section.
Equivalent so called t-SNE visualizations are shown in the SI, Figs.~S4-S7.

\subsection*{Exploring the material space}\label{subSecExplore}

\begin{figure*}[ht]
    \centering
    \includegraphics[width=0.85\textwidth]{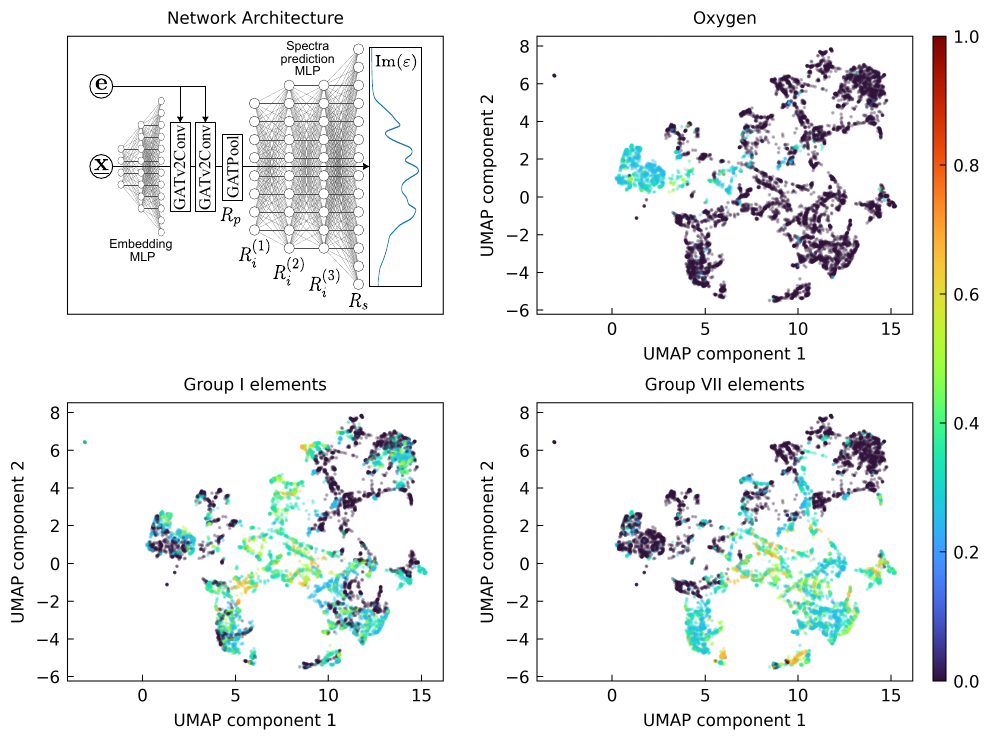}
    \caption{
    (Top left) Structure of the \textsc{OptiMate} GAT consisting of (from left to right) an atom embedding MLP learning an adapted periodic system, a varying number of message passing blocks allowing for communication between atoms, and a spectra prediction MLP, see Ref.~\onlinecite{Grunert2024b}.
    The high dimensional representations $R$ (vectors) throughout the model are highlighted and labeled accordingly. 
    $R_p$ are the latent embeddings after the pooling operation, $R_i^{(n)}$ are the internal embeddings of the spectra prediction MLP and $R_s$ are the output of the last layer of the spectra prediction MLP, i.e., the materials optical spectrum. 
    The remaining subplots show UMAPs of the latent embeddings after the pooling operation, i.e., of the vector $R_p$, colored according to the prevalence of certain elements: 
    (Top right) oxygen, (Bottom left) group I elements, (Bottom right) group VII elements.
    The colorbar indicates the percentage of all sites in each material that belong to the given element.
    }
    \label{fig:UMAP_elems}
\end{figure*}

We begin by focusing on $R_p$, i.e., the latent embeddings after the pooling step, when the model first establishes a single high-dimensional representation of fixed length for each material, as required for the UMAP algorithm.
For added clarity, we note that before the pooling step, \textsc{OptiMate} acts on individual atoms and their neighbours.
Thus there is no obvious way to consistently extract a vector of fixed length per material independent of the number of inequivalent atoms involved.
Instead, one might carry out a UMAP of all sites in each material and thus obtain a lower-dimensional representation of each chemical environment.
The resulting low-dimensional embeddings are visualized in Fig.~\ref{fig:UMAP_elems}, colored according to the prevalence of certain elements (other elements are shown in the SI, Figs.~S2 and S3, as well as an interactive version of all UMAP plots shown in this work).
As can be clearly seen, the model clusters according to the elements present in each compound, which allows us to address the question raised above, namely whether \textsc{OptiMate} learns which elements are more important and which elements are less important for the optical properties.
As expected, the importance varies from element to element:
For example, most oxygen-containing materials are placed in one large cluster of oxoanions with a few smaller clusters elsewhere; see Fig.~\ref{fig:UMAP_elems} (top right).
Compounds containing nitrogen or carbon show a similar behavior, see Fig.~S3 of the SI.
A cluster containing compounds with nitrite and nitrate anions is positioned between them (see overlap between Fig.~\ref{fig:UMAP_elems} (top right) and Fig.~S3 (left), close to a cluster containing compounds with carbonate (see overlap between Fig.~\ref{fig:UMAP_elems} (top right) and Fig.~S3 (right)).
Similar behavior can be observed for many other anions.
Compounds containing group I and group VII elements (usually cations) on the other hand are distributed all over the material space.
A cursory analysis suggests that these elements mostly serve as point-charge-like "counter" or "filler" ions to the elements and anions dominating the compound's optical response.
For example, materials containing phosphate ions are placed close together irregardless of their "cation".
Why the model seems to give more weight to the anion in these cases is not immediately clear, but is mostly consistent with chemical and physical intuition as well as experience:
In the standard textbook example of III-V and II-VI semiconductors, the valence band is derived  from the anion while the conduction band originates from the cation, e.g., GaAs and ZnSe.\cite{Pashley1989}
In a slightly simplified picture, the optical spectra are dominated by valence-band-to-conduction-band transitions, and one might thus expect equal influence of the anion and cation.
However, the simple textbook description is not entirely correct, since the anion often also contributes significantly to the conduction band (the contribution being defined by the partial density of states).
We have also recently observed a stronger influence of the valence band on the exciton binding energy,\cite{Grunert2024a} another closely related optical property.
One might therefore be inclined to understand the model's emphasis on the anion
-- and even hope that \textsc{OptiMate} and its successors will enable similar, truly new observations.  

As an invitation to own explorations, we mention two more observations:
As one might expect, the model places the noble gas solids separately from the other materials (i.e., see the dots around UMAP$_1 = 2$, UMAP$_2 = 1$ in the UMAPs of Fig.~\ref{fig:UMAP_elems}).
Interestingly, for reasons we do not (yet) understand and which will be investigated in the future, the model places all compounds containing boron hydride in the top left-hand corner, separate from the rest of the material space.
Closer examination of the internal structure of individual clusters reveals that many materials that a trained materials scientist would classify as similar, at least in terms of their optical properties, are in fact very close together.
For example, one subcluster in the "general group V anion region" contains all III-V semiconductors, while another subcluster in the "general group VI ion region" contains all compounds with an AB$_3$ ion, where A is a group V element and B is a group VI element.
This suggests that the model more or less recovers the basic chemistry, but not just for (relatively) few classes of materials, as a human chemist would do, but for the whole material space -- arguably one could say that the model acts like a "room full of chemists". 
We expect a lot of research to be conducted on observations and interpretations like those reported here and in the SI. 

\begin{figure*}[ht]
    \centering
    \includegraphics[width=0.85\textwidth]{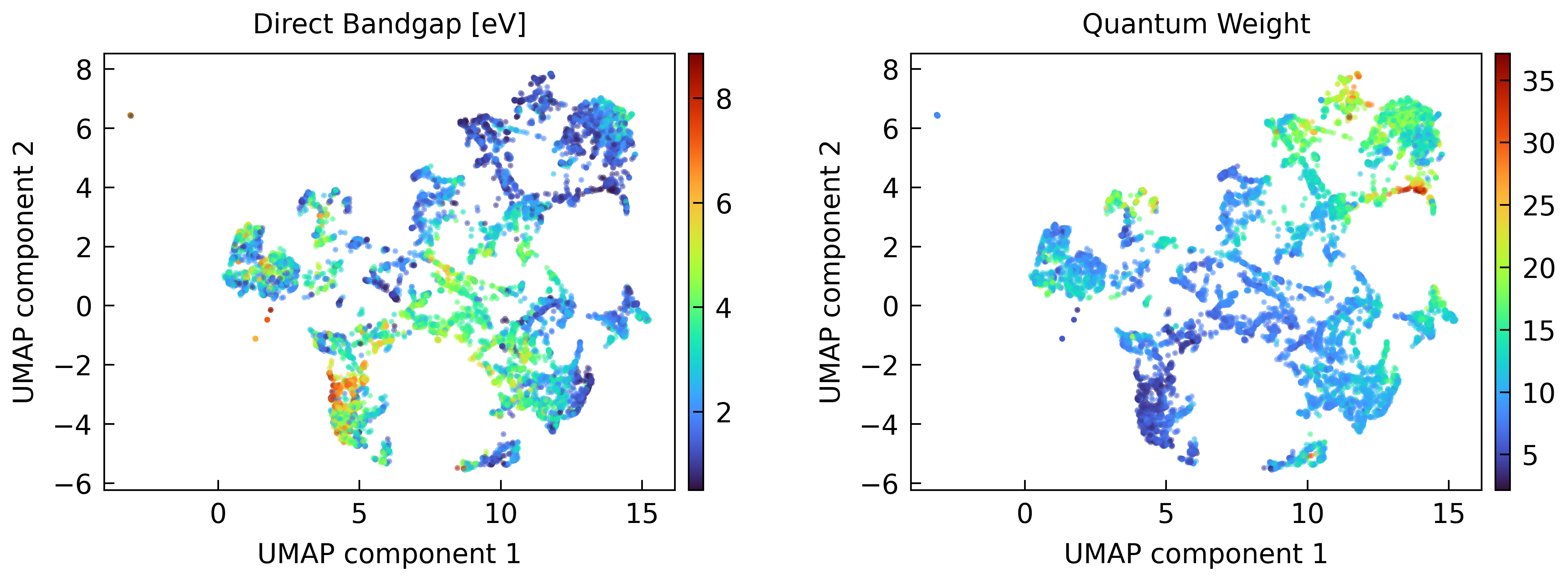}
    \caption{
    UMAP of the latent embeddings after the pooling operation $R_p$, colored according to (left) the direct band gap in eV as obtained from DFT and (right) the quantum weight calculated from the predicted spectra.
    }
    \label{fig:UMAP_optics}
\end{figure*}

Since the model is designed and trained to accurately predict optical spectra, it is an obvious idea to see how the dimensionality-reduced embeddings of their materials relate to their optical spectra.
The model was trained on entire spectrum from $0$ to $20$~eV. 
Thus, one should not expect excellent clustering with respect to optical properties favored by humans, such as the direct bandgap. 
Indeed, a significant variation inside clusters can be observed in the left panel of Fig.~\ref{fig:UMAP_optics}.
However, the spectra of nearby materials generally show a strong visual similarity, of which the reader can easily convince herself or himself with the help of the interactive version (see the SI).
To analyze this observation quantitatively, we need a property that classifies the entire spectrum.
For this, we use the "quantum weight" from Ref.~\onlinecite{Hung2024}, i.e., the integral of the spectra over the calculated region.
The "quantum weight" itself is motivated by the Thomas-Reiche-Kuhn or f-sum rule. 
The resulting clustering is shown in Fig.~\ref{fig:UMAP_optics}, and indeed, a reasonably smooth distribution is evident for a large part of the material space.

\subsection*{Transforming the material space}\label{subSecTransform}

\begin{figure*}[ht]
    \centering
    \includegraphics[width=0.85\textwidth]{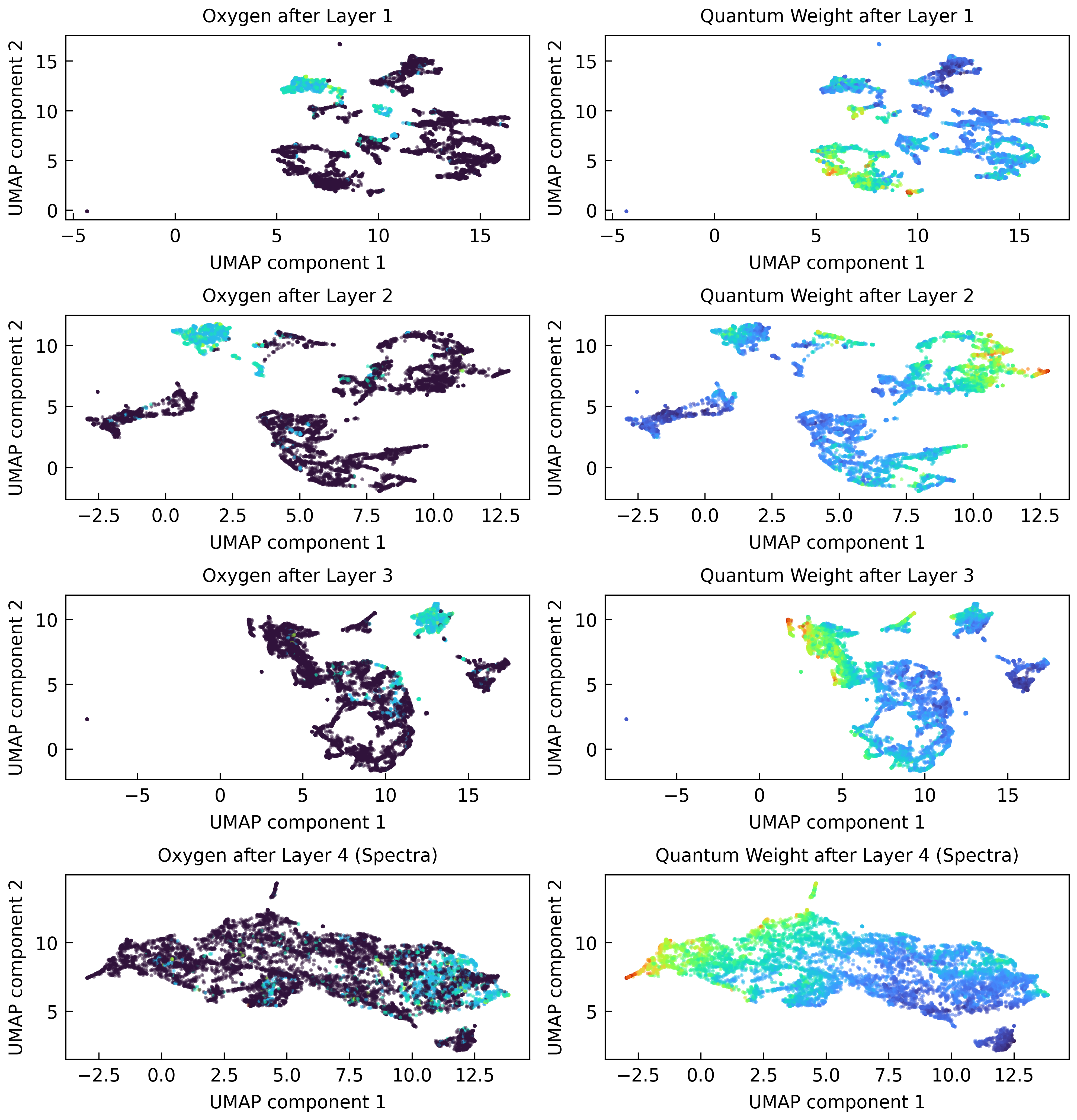}
    \caption{
    UMAP of the latent embeddings after, from top to bottom, the first, second, and third  layer of the spectra prediction MLP $R_i^{(1-3)}$ and the final output of the spectra prediction MLP $R_s$, colored according to (left) the share of oxygen in the compounds and (right) the quantum weight calculated from the predicted spectra.
    }
    \label{fig:UMAP_mlp}
\end{figure*}

We now turn to investigating how the spectra prediction MLP processes the embedding obtained after pooling via its three hidden layers to finally obtain the predicted spectra, specifically how the embeddings $R_p$ discussed above are transformed through the internal embeddings $R_i^{(1-3)}$ to the predicted spectra $R_s$.
This question is of less interest from a chemistry or materials science point of view, but interesting from the machine learning point of view. 
The resulting UMAPs, colored according to the oxygen content and the quantum weight, are shown in Fig.~\ref{fig:UMAP_mlp}.
Obviously, the rich chemistry of the materials space seen at the level of $R_p$ is transformed layer by layer into the more continuous picture of spectra $R_s$ which --  very roughly speaking -- as functions of photon energy all start with vanishing values below the band gap, rise to a distinct peak and then slowly fall off.
This is reflected in the UMAPS, as clusters join together to form larger clusters when going from $R_p$ through $R_i^{(1-3)}$ to $R_s$.
The three separate clusters after layer 3 contain predominantly compounds of the elements with the highest electronegativity on the Allred-Rochow scale, namely fluorine, oxygen, and nitrogen (from right to left).

A more detailed discussion of these results, especially the formation of individual clusters and their relationship to each other, is certainly interesting, but we defer to the SI, which includes an interactive visualization of the associated UMAP plots. 
We encourage all readers to explore the material space for themselves, which is made possible by our interactive UMAP plots.
We now present another, more directly applicable use of the learned material space.

\subsection*{Use case: Critical and rare materials}\label{subSecUse}

In recent decades, technologies based on semiconductors and insulators have proliferated around the world, leading to an explosion in resource consumption.\cite{Schrijvers2020}    
Many of the incorporated elements are problematic in one sense or another\cite{Nuss2014} -- some of the elements used are rare, e.g., indium, which is obtained primarily as a byproduct of zinc production \cite{Gunn2013-uc}, others are toxic, e.g., arsenic.\cite{Ratnaike2003} 
As a result, government agencies, the research community, and major industries are actively searching for alternative materials.
After establishing that the model creates a meaningful representation of the (optical) material space, we show how this learned material space can be used for the purpose of identifying alternative materials. 

\begin{figure*}[ht]
    \centering
    \includegraphics[width=0.85\textwidth]{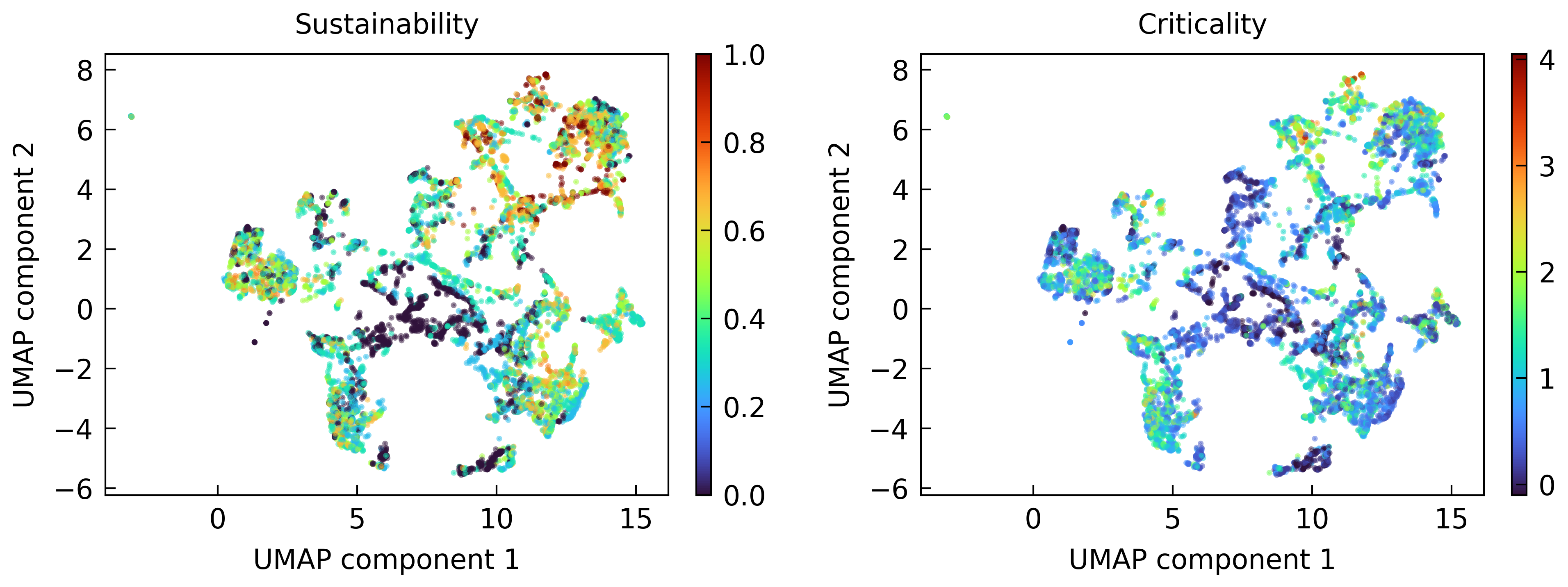}
    \caption{
    UMAP of the latent embeddings after the pooling operation, colored according to (left) the sustainability and (right) the criticality.
    Higher values correspond to less sustainable/more critical materials.
    }
    \label{fig:UMAP_sust}
\end{figure*}

There are many different quantifications of criticality and the importance of finding alternatives for certain materials (see e.g., Ref.~\onlinecite{Glavi2007} for a discussion of sustainability terms).
Here we focus on two criteria, sustainability as defined by the European Chemical Society, \cite{euchemsElementScarcity} which measures the relative availability of individual elements and whether or not the current global consumption rate is sustainable based on reserves, and criticality as defined by the European Union,\cite{EuCrit} which measures the importance and supply risk of various elements (some in the form of ores, see SI) for certain key technologies.
We note that even these two narrow definitions of these criteria result in relatively frequently changing results -- the European Chemical Society updates its assessment annually, and the European Union updates its report every 3 years.
We define the 'sustainability of a compound' based on the average sustainability of its constituent elements, weighted by their abundance in the material, and extract the sustainability of an element from Ref.~\onlinecite{euchemsElementScarcity} by treating it as a binary variable, i.e., an element is either sufficiently 'sustainable' (green in Ref.~\onlinecite{euchemsElementScarcity}) or not (orange/red).
We define the 'criticality of a compound' analogously, using the numerical values for supply chain risk given in Ref.~\onlinecite{EuCrit}.
A visualization of the resulting sustainability and criticality scores can be found in the SI.
The resulting diagrams are shown in Fig.~\ref{fig:UMAP_sust}.
A base level of clustering can still be observed. 
This is to be expected, as it was shown above that the model clusters materials according to chemical rules. However, the sustainability or criticality of a material are also related - in a more-or-less complex fashion - to chemical rules.
We again provide interactive versions of these maps in the SI. 

\begin{figure}[ht]
    \centering
    \includegraphics[width=0.65\textwidth]{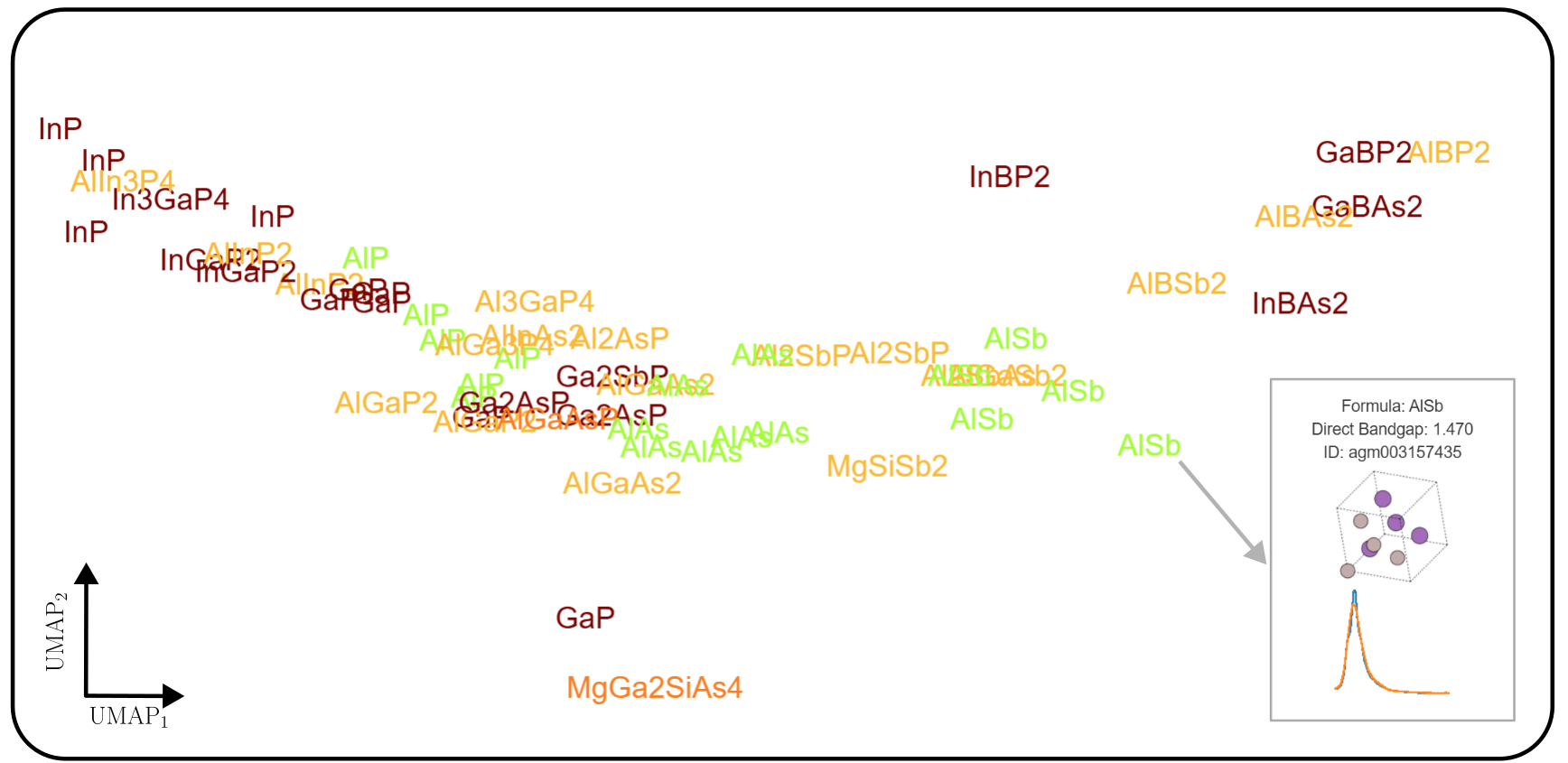}
    \caption{
    Screenshot of the interactive UMAP colored by sustainability (see SI), zoomed to the region of III-V semiconductor compounds.
    The grid lines and axis have been removed for visual clarity.
    The interactive UMAPs show the name of the compound at their respective position.
    When hovering over a compound (here: AlSb), a tooltip is shown, containing the direct band gap, the ID of the material in the Alexandria database, \cite{Schmidt2024} a visualization of the crystal structure and the \textit{ab initio} calculated spectra (blue) and the spectra predicted by \textsc{OptiMate} (orange).
    }
    \label{fig:Inset}
\end{figure}

At this point, we would like to highlight one region of the map in particular, which we discuss as an example of how this map of the material space can be used to find more sustainable or less critical materials.
We take a closer look at the III-V compounds used in state-of-the-art multijunction photovoltaic (PV) cells \cite{Hannappel2024} and various other optoelectronic applications.
A close-up of this region of the map is shown in Fig.~\ref{fig:Inset}.
As mentioned above, III-V materials are packed close together and cluster more according to their anion, i.e., phosphorus (P), arsenic (As), etc.
Indium (In) and gallium (Ga) are considered to be unsustainable materials,\cite{euchemsElementScarcity} with Ga also being considered a highly critical material by the European Union, and there have been ongoing research efforts to reduce their use, for example in the industrially relevant materials indium phosphide (InP) and gallium phosphide (GaP).
A primary thrust in the PV community is to (partially) replace the critical cations with the non-critical cation aluminum (Al), and indeed we can recover the same conclusion from the learned material space: 
For example, modifications of the more sustainable aluminum indium phosphide (AlInP$_2$, AlIn$_3$P$_4$) are very close to modifications of the unsustainable InP and GaP, but further away from optically less similar materials such as aluminum antimonide (AlSb).
A little further away, the very uncritical hypothetical material SiB$_2$S is found to have very similar optical properties to the III-V materials and thus could be a promising alternative.
At this point, we can only encourage the reader to search for possible alternatives to other materials in our interactive UMAP plots in the SI.
We also note that not only In and Ga are targets for substitution due to their criticality, but the use of arsenic-containing III-V compounds is also controversial due to their toxicity.
A similar analysis as done here for criticality and sustainability could equally well be done for toxicity, and indeed for any other property that can be quantified, e.g., estimated production cost.
We note that while in this work we focus on reduction to two dimensions for visualization, one can also directly identify similar materials in the higher dimensional space at the cost of human interpretability. 

\section{Discussion}
The results presented in this work are twofold:
First, we have shown that modern machine learning models can construct a meaningful internal representation of the material space, similar to that of trained materials scientists.
This is obviously a very intriguing result, as it explains the good performance of these methods, and it is reassuring to see such a close agreement between human and machine intelligence.
However, this does not mean that current machine learning models can be fully trusted -- it is still possible that the model will repeatedly put exotic materials in the "wrong position", much like an overconfident student in an exam (for example, the model with original weights from Ref.~\onlinecite{Grunert2024b} is certain that layered and three-dimensional boron nitrides, which are both in the test set, are extremely similar).
Second, we have shown how this internal material space can be extracted, explored, and used for relevant technological applications, e.g., for identifying more sustainable and less critical materials for photovoltaics.
The presented strategy is applicable for both a wide range of materials and a wide range of technological applications, given that there is enough data to train a machine learning model to construct sensible internal representations.
With the current pace of data generation efforts in many different fields of materials science,\cite{Jain2013,Schmidt2024,Barroso-Luque2024} we believe that this will not be a problem for much longer.
In the future, the analysis presented here should be applied to other machine learning models, either trained on other technologically important properties, such as the transition temperature of superconductors,\cite{Cerqueira2023} or applied to multimodal and foundation models, trained on a variety of properties simultaneously.
In the latter case, a recovery of the entire material space (and not just the "projection" in terms of optical properties) could be the result. 
Finally, we believe that comparing the learned internal representations between different architectures will offer valuable insight while also potentially opening another dimension of model performance beyond simply evaluating the accuracy on a test set.

\section{Methods}
The model architecture and the hyperparameters are identical to the \textsc{OptiMate} model for the prediction of $\mathrm{Tr}[\mathrm{Im}(\varepsilon_{ij})]/3$ with a broadening of $300$~meV from Ref.~\onlinecite{Grunert2024b}.
The analyzed dataset is also taken from Ref.~\onlinecite{Grunert2024b}.
As mentioned in the main text, instead of performing a traditional training-validation-test split on the database, we train the model on the entire dataset of $\mathrm{Tr}[\mathrm{Im}(\varepsilon_{ij})]/3$ calculated with a broadening of $300$~meV.

After training, we extract latent embeddings for all materials after the GATPool operation and all layers of the spectra-predicting MLP, with the output fourth layer being the predicted spectrum.
As the model is to a very small extent non-deterministic, the latent embeddings are saved to ensure consistency between runs.

UMAPs are then performed on the latent embeddings to reduce the high-dimensional embeddings to visualizable two-dimensional values using the following parameters:
Number of neighbours: 30;
Minimum distance: 0.01;
Metric: Euclidean.
We note that varying the parameters (within a reasonable range) changes the visual appearance of the resulting UMAP plots, but does not change the conclusions drawn.
To ensure reproducibility, all UMAPs are performed with a fixed random seed, namely 0.
Small variations are still possible when switching the operating system due to variations in the implementation of pseudorandom numbers. 
The images shown in this work were generated using the Linux distribution Debian.

\section{Data and Code Availability}
The data and code supporting the findings of this study will be made available upon publication and have been attached as additional material for the reviewers.

\section{Acknowledgments}
The authors thank the staff of the Compute Center of the Technische Universität Ilmenau and especially Mr.~Henning~Schwanbeck for providing an excellent research environment. 
This work is supported by the Deutsche Forschungsgemeinschaft DFG (Project 537033066) and the Project SustEnMat of the Carl Zeiss Stiftung (Funding code: P2023-02-008).

\section{Competing interests}
The authors declare no competing interests.

\providecommand{\noopsort}[1]{}\providecommand{\singleletter}[1]{#1}%


\begin{thebibliography}{41}%
\makeatletter
\providecommand \@ifxundefined [1]{%
 \@ifx{#1\undefined}
}%
\providecommand \@ifnum [1]{%
 \ifnum #1\expandafter \@firstoftwo
 \else \expandafter \@secondoftwo
 \fi
}%
\providecommand \@ifx [1]{%
 \ifx #1\expandafter \@firstoftwo
 \else \expandafter \@secondoftwo
 \fi
}%
\providecommand \natexlab [1]{#1}%
\providecommand \enquote  [1]{``#1''}%
\providecommand \bibnamefont  [1]{#1}%
\providecommand \bibfnamefont [1]{#1}%
\providecommand \citenamefont [1]{#1}%
\providecommand \href@noop [0]{\@secondoftwo}%
\providecommand \href [0]{\begingroup \@sanitize@url \@href}%
\providecommand \@href[1]{\@@startlink{#1}\@@href}%
\providecommand \@@href[1]{\endgroup#1\@@endlink}%
\providecommand \@sanitize@url [0]{\catcode `\\12\catcode `\$12\catcode `\&12\catcode `\#12\catcode `\^12\catcode `\_12\catcode `\%12\relax}%
\providecommand \@@startlink[1]{}%
\providecommand \@@endlink[0]{}%
\providecommand \url  [0]{\begingroup\@sanitize@url \@url }%
\providecommand \@url [1]{\endgroup\@href {#1}{\urlprefix }}%
\providecommand \urlprefix  [0]{URL }%
\providecommand \Eprint [0]{\href }%
\providecommand \doibase [0]{https://doi.org/}%
\providecommand \selectlanguage [0]{\@gobble}%
\providecommand \bibinfo  [0]{\@secondoftwo}%
\providecommand \bibfield  [0]{\@secondoftwo}%
\providecommand \translation [1]{[#1]}%
\providecommand \BibitemOpen [0]{}%
\providecommand \bibitemStop [0]{}%
\providecommand \bibitemNoStop [0]{.\EOS\space}%
\providecommand \EOS [0]{\spacefactor3000\relax}%
\providecommand \BibitemShut  [1]{\csname bibitem#1\endcsname}%
\let\auto@bib@innerbib\@empty
\bibitem [{\citenamefont {Batatia}\ \emph {et~al.}(2022{\natexlab{a}})\citenamefont {Batatia}, \citenamefont {Batzner}, \citenamefont {Kov{\'a}cs}, \citenamefont {Musaelian}, \citenamefont {Simm}, \citenamefont {Drautz}, \citenamefont {Ortner}, \citenamefont {Kozinsky},\ and\ \citenamefont {Cs{\'a}nyi}}]{Batatia2022Design}%
  \BibitemOpen
  \bibfield  {author} {\bibinfo {author} {\bibfnamefont {I.}~\bibnamefont {Batatia}}, \bibinfo {author} {\bibfnamefont {S.}~\bibnamefont {Batzner}}, \bibinfo {author} {\bibfnamefont {D.~P.}\ \bibnamefont {Kov{\'a}cs}}, \bibinfo {author} {\bibfnamefont {A.}~\bibnamefont {Musaelian}}, \bibinfo {author} {\bibfnamefont {G.~N.~C.}\ \bibnamefont {Simm}}, \bibinfo {author} {\bibfnamefont {R.}~\bibnamefont {Drautz}}, \bibinfo {author} {\bibfnamefont {C.}~\bibnamefont {Ortner}}, \bibinfo {author} {\bibfnamefont {B.}~\bibnamefont {Kozinsky}},\ and\ \bibinfo {author} {\bibfnamefont {G.}~\bibnamefont {Cs{\'a}nyi}},\ }\href {https://doi.org/10.48550/arXiv.2205.06643} {\enquote {\bibinfo {title} {The {D}esign {S}pace of {E(3)}-{E}quivariant {A}tom-{C}entered {I}nteratomic {P}otentials},}\ } (\bibinfo {year} {2022}{\natexlab{a}}),\ \Eprint {https://arxiv.org/abs/2205.06643} {arXiv:2205.06643} \BibitemShut {NoStop}%
\bibitem [{\citenamefont {Batatia}\ \emph {et~al.}(2022{\natexlab{b}})\citenamefont {Batatia}, \citenamefont {Kovacs}, \citenamefont {Simm}, \citenamefont {Ortner},\ and\ \citenamefont {Csanyi}}]{Batatia2022mace}%
  \BibitemOpen
  \bibfield  {author} {\bibinfo {author} {\bibfnamefont {I.}~\bibnamefont {Batatia}}, \bibinfo {author} {\bibfnamefont {D.~P.}\ \bibnamefont {Kovacs}}, \bibinfo {author} {\bibfnamefont {G.~N.~C.}\ \bibnamefont {Simm}}, \bibinfo {author} {\bibfnamefont {C.}~\bibnamefont {Ortner}},\ and\ \bibinfo {author} {\bibfnamefont {G.}~\bibnamefont {Csanyi}},\ }\bibfield  {title} {\enquote {\bibinfo {title} {{MACE}: {H}igher {O}rder {E}quivariant {M}essage {P}assing {N}eural {N}etworks for {F}ast and {A}ccurate {F}orce {F}ields},}\ }in\ \href {https://openreview.net/forum?id=YPpSngE-ZU} {\emph {\bibinfo {booktitle} {Advances in Neural Information Processing Systems}}},\ \bibinfo {editor} {edited by\ \bibinfo {editor} {\bibfnamefont {A.~H.}\ \bibnamefont {Oh}}, \bibinfo {editor} {\bibfnamefont {A.}~\bibnamefont {Agarwal}}, \bibinfo {editor} {\bibfnamefont {D.}~\bibnamefont {Belgrave}},\ and\ \bibinfo {editor} {\bibfnamefont {K.}~\bibnamefont {Cho}}}\ (\bibinfo {year} {2022})\BibitemShut {NoStop}%
\bibitem [{\citenamefont {Yang}\ \emph {et~al.}(2024)\citenamefont {Yang}, \citenamefont {Hu}, \citenamefont {Zhou}, \citenamefont {Liu}, \citenamefont {Shi}, \citenamefont {Li}, \citenamefont {Li}, \citenamefont {Chen}, \citenamefont {Chen}, \citenamefont {Zeni}, \citenamefont {Horton}, \citenamefont {Pinsler}, \citenamefont {Fowler}, \citenamefont {Z\"{u}gner}, \citenamefont {Xie}, \citenamefont {Smith}, \citenamefont {Sun}, \citenamefont {Wang}, \citenamefont {Kong}, \citenamefont {Liu}, \citenamefont {Hao},\ and\ \citenamefont {Lu}}]{MatterSim2024}%
  \BibitemOpen
  \bibfield  {author} {\bibinfo {author} {\bibfnamefont {H.}~\bibnamefont {Yang}}, \bibinfo {author} {\bibfnamefont {C.}~\bibnamefont {Hu}}, \bibinfo {author} {\bibfnamefont {Y.}~\bibnamefont {Zhou}}, \bibinfo {author} {\bibfnamefont {X.}~\bibnamefont {Liu}}, \bibinfo {author} {\bibfnamefont {Y.}~\bibnamefont {Shi}}, \bibinfo {author} {\bibfnamefont {J.}~\bibnamefont {Li}}, \bibinfo {author} {\bibfnamefont {G.}~\bibnamefont {Li}}, \bibinfo {author} {\bibfnamefont {Z.}~\bibnamefont {Chen}}, \bibinfo {author} {\bibfnamefont {S.}~\bibnamefont {Chen}}, \bibinfo {author} {\bibfnamefont {C.}~\bibnamefont {Zeni}}, \bibinfo {author} {\bibfnamefont {M.}~\bibnamefont {Horton}}, \bibinfo {author} {\bibfnamefont {R.}~\bibnamefont {Pinsler}}, \bibinfo {author} {\bibfnamefont {A.}~\bibnamefont {Fowler}}, \bibinfo {author} {\bibfnamefont {D.}~\bibnamefont {Z\"{u}gner}}, \bibinfo {author} {\bibfnamefont {T.}~\bibnamefont {Xie}}, \bibinfo {author} {\bibfnamefont {J.}~\bibnamefont {Smith}}, \bibinfo {author} {\bibfnamefont
  {L.}~\bibnamefont {Sun}}, \bibinfo {author} {\bibfnamefont {Q.}~\bibnamefont {Wang}}, \bibinfo {author} {\bibfnamefont {L.}~\bibnamefont {Kong}}, \bibinfo {author} {\bibfnamefont {C.}~\bibnamefont {Liu}}, \bibinfo {author} {\bibfnamefont {H.}~\bibnamefont {Hao}},\ and\ \bibinfo {author} {\bibfnamefont {Z.}~\bibnamefont {Lu}},\ }\href {https://doi.org/10.48550/ARXIV.2405.04967} {\enquote {\bibinfo {title} {Matter{S}im: {A} {D}eep {L}earning {A}tomistic {M}odel {A}cross {E}lements, {T}emperatures and {P}ressures},}\ } (\bibinfo {year} {2024}),\ \Eprint {https://arxiv.org/abs/2405.04967} {arXiv:2405.04967} \BibitemShut {NoStop}%
\bibitem [{\citenamefont {Li}\ \emph {et~al.}(2022)\citenamefont {Li}, \citenamefont {Wang}, \citenamefont {Zou}, \citenamefont {Ye}, \citenamefont {Xu}, \citenamefont {Gong}, \citenamefont {Duan},\ and\ \citenamefont {Xu}}]{Li2022}%
  \BibitemOpen
  \bibfield  {author} {\bibinfo {author} {\bibfnamefont {H.}~\bibnamefont {Li}}, \bibinfo {author} {\bibfnamefont {Z.}~\bibnamefont {Wang}}, \bibinfo {author} {\bibfnamefont {N.}~\bibnamefont {Zou}}, \bibinfo {author} {\bibfnamefont {M.}~\bibnamefont {Ye}}, \bibinfo {author} {\bibfnamefont {R.}~\bibnamefont {Xu}}, \bibinfo {author} {\bibfnamefont {X.}~\bibnamefont {Gong}}, \bibinfo {author} {\bibfnamefont {W.}~\bibnamefont {Duan}},\ and\ \bibinfo {author} {\bibfnamefont {Y.}~\bibnamefont {Xu}},\ }\bibfield  {title} {\enquote {\bibinfo {title} {Deep-learning density functional theory {H}amiltonian for efficient ab initio electronic-structure calculation},}\ }\href {https://doi.org/10.1038/s43588-022-00265-6} {\bibfield  {journal} {\bibinfo  {journal} {Nat. Comput. Sci.}\ }\textbf {\bibinfo {volume} {2}},\ \bibinfo {pages} {367–377} (\bibinfo {year} {2022})}\BibitemShut {NoStop}%
\bibitem [{\citenamefont {Gong}\ \emph {et~al.}(2023)\citenamefont {Gong}, \citenamefont {Li}, \citenamefont {Zou}, \citenamefont {Xu}, \citenamefont {Duan},\ and\ \citenamefont {Xu}}]{Gong2023}%
  \BibitemOpen
  \bibfield  {author} {\bibinfo {author} {\bibfnamefont {X.}~\bibnamefont {Gong}}, \bibinfo {author} {\bibfnamefont {H.}~\bibnamefont {Li}}, \bibinfo {author} {\bibfnamefont {N.}~\bibnamefont {Zou}}, \bibinfo {author} {\bibfnamefont {R.}~\bibnamefont {Xu}}, \bibinfo {author} {\bibfnamefont {W.}~\bibnamefont {Duan}},\ and\ \bibinfo {author} {\bibfnamefont {Y.}~\bibnamefont {Xu}},\ }\bibfield  {title} {\enquote {\bibinfo {title} {General framework for {E}(3)-equivariant neural network representation of density functional theory {H}amiltonian},}\ }\href {https://doi.org/10.1038/s41467-023-38468-8} {\bibfield  {journal} {\bibinfo  {journal} {Nat. Commun.}\ }\textbf {\bibinfo {volume} {14}} (\bibinfo {year} {2023}),\ 10.1038/s41467-023-38468-8}\BibitemShut {NoStop}%
\bibitem [{\citenamefont {Tang}\ \emph {et~al.}(2024)\citenamefont {Tang}, \citenamefont {Li}, \citenamefont {Lin}, \citenamefont {Gong}, \citenamefont {Jin}, \citenamefont {He}, \citenamefont {Jiang}, \citenamefont {Ren}, \citenamefont {Duan},\ and\ \citenamefont {Xu}}]{Tang2024}%
  \BibitemOpen
  \bibfield  {author} {\bibinfo {author} {\bibfnamefont {Z.}~\bibnamefont {Tang}}, \bibinfo {author} {\bibfnamefont {H.}~\bibnamefont {Li}}, \bibinfo {author} {\bibfnamefont {P.}~\bibnamefont {Lin}}, \bibinfo {author} {\bibfnamefont {X.}~\bibnamefont {Gong}}, \bibinfo {author} {\bibfnamefont {G.}~\bibnamefont {Jin}}, \bibinfo {author} {\bibfnamefont {L.}~\bibnamefont {He}}, \bibinfo {author} {\bibfnamefont {H.}~\bibnamefont {Jiang}}, \bibinfo {author} {\bibfnamefont {X.}~\bibnamefont {Ren}}, \bibinfo {author} {\bibfnamefont {W.}~\bibnamefont {Duan}},\ and\ \bibinfo {author} {\bibfnamefont {Y.}~\bibnamefont {Xu}},\ }\bibfield  {title} {\enquote {\bibinfo {title} {A deep equivariant neural network approach for efficient hybrid density functional calculations},}\ }\href {https://doi.org/10.1038/s41467-024-53028-4} {\bibfield  {journal} {\bibinfo  {journal} {Nat. Commun.}\ }\textbf {\bibinfo {volume} {15}} (\bibinfo {year} {2024}),\ 10.1038/s41467-024-53028-4}\BibitemShut {NoStop}%
\bibitem [{\citenamefont {Pilania}\ \emph {et~al.}(2016)\citenamefont {Pilania}, \citenamefont {Mannodi-Kanakkithodi}, \citenamefont {Uberuaga}, \citenamefont {Ramprasad}, \citenamefont {Gubernatis},\ and\ \citenamefont {Lookman}}]{Pilania2016}%
  \BibitemOpen
  \bibfield  {author} {\bibinfo {author} {\bibfnamefont {G.}~\bibnamefont {Pilania}}, \bibinfo {author} {\bibfnamefont {A.}~\bibnamefont {Mannodi-Kanakkithodi}}, \bibinfo {author} {\bibfnamefont {B.~P.}\ \bibnamefont {Uberuaga}}, \bibinfo {author} {\bibfnamefont {R.}~\bibnamefont {Ramprasad}}, \bibinfo {author} {\bibfnamefont {J.~E.}\ \bibnamefont {Gubernatis}},\ and\ \bibinfo {author} {\bibfnamefont {T.}~\bibnamefont {Lookman}},\ }\bibfield  {title} {\enquote {\bibinfo {title} {Machine learning bandgaps of double perovskites},}\ }\href {https://doi.org/10.1038/srep19375} {\bibfield  {journal} {\bibinfo  {journal} {Sci. Rep.}\ }\textbf {\bibinfo {volume} {6}},\ \bibinfo {pages} {6} (\bibinfo {year} {2016})}\BibitemShut {NoStop}%
\bibitem [{\citenamefont {Lee}\ \emph {et~al.}(2016)\citenamefont {Lee}, \citenamefont {Seko}, \citenamefont {Shitara}, \citenamefont {Nakayama},\ and\ \citenamefont {Tanaka}}]{Lee2016}%
  \BibitemOpen
  \bibfield  {author} {\bibinfo {author} {\bibfnamefont {J.}~\bibnamefont {Lee}}, \bibinfo {author} {\bibfnamefont {A.}~\bibnamefont {Seko}}, \bibinfo {author} {\bibfnamefont {K.}~\bibnamefont {Shitara}}, \bibinfo {author} {\bibfnamefont {K.}~\bibnamefont {Nakayama}},\ and\ \bibinfo {author} {\bibfnamefont {I.}~\bibnamefont {Tanaka}},\ }\bibfield  {title} {\enquote {\bibinfo {title} {Prediction model of band gap for inorganic compounds by combination of density functional theory calculations and machine learning techniques},}\ }\href {https://doi.org/10.1103/PhysRevB.93.115104} {\bibfield  {journal} {\bibinfo  {journal} {Phys. Rev. B}\ }\textbf {\bibinfo {volume} {93}},\ \bibinfo {pages} {115104} (\bibinfo {year} {2016})}\BibitemShut {NoStop}%
\bibitem [{\citenamefont {Xie}\ and\ \citenamefont {Grossman}(2018)}]{Xie2018}%
  \BibitemOpen
  \bibfield  {author} {\bibinfo {author} {\bibfnamefont {T.}~\bibnamefont {Xie}}\ and\ \bibinfo {author} {\bibfnamefont {J.~C.}\ \bibnamefont {Grossman}},\ }\bibfield  {title} {\enquote {\bibinfo {title} {Crystal {G}raph {C}onvolutional {N}eural {N}etworks for an {A}ccurate and {I}nterpretable {P}rediction of {M}aterial {P}roperties},}\ }\href {https://doi.org/10.1103/PhysRevLett.120.145301} {\bibfield  {journal} {\bibinfo  {journal} {Phys. Rev. Lett.}\ }\textbf {\bibinfo {volume} {120}},\ \bibinfo {pages} {145301} (\bibinfo {year} {2018})}\BibitemShut {NoStop}%
\bibitem [{\citenamefont {Rajan}\ \emph {et~al.}(2018)\citenamefont {Rajan}, \citenamefont {Mishra}, \citenamefont {Satsangi}, \citenamefont {Vaish}, \citenamefont {Mizuseki}, \citenamefont {Lee},\ and\ \citenamefont {Singh}}]{Rajan2018}%
  \BibitemOpen
  \bibfield  {author} {\bibinfo {author} {\bibfnamefont {A.~C.}\ \bibnamefont {Rajan}}, \bibinfo {author} {\bibfnamefont {A.}~\bibnamefont {Mishra}}, \bibinfo {author} {\bibfnamefont {S.}~\bibnamefont {Satsangi}}, \bibinfo {author} {\bibfnamefont {R.}~\bibnamefont {Vaish}}, \bibinfo {author} {\bibfnamefont {H.}~\bibnamefont {Mizuseki}}, \bibinfo {author} {\bibfnamefont {K.-R.}\ \bibnamefont {Lee}},\ and\ \bibinfo {author} {\bibfnamefont {A.~K.}\ \bibnamefont {Singh}},\ }\bibfield  {title} {\enquote {\bibinfo {title} {Machine-{L}earning-{A}ssisted {A}ccurate {B}and {G}ap {P}redictions of {F}unctionalized {MX}ene},}\ }\href {https://doi.org/10.1021/acs.chemmater.8b00686} {\bibfield  {journal} {\bibinfo  {journal} {Chem. Mater.}\ }\textbf {\bibinfo {volume} {30}},\ \bibinfo {pages} {4031–4038} (\bibinfo {year} {2018})}\BibitemShut {NoStop}%
\bibitem [{\citenamefont {Zhuo}, \citenamefont {Mansouri~Tehrani},\ and\ \citenamefont {Brgoch}(2018)}]{Zhuo2018}%
  \BibitemOpen
  \bibfield  {author} {\bibinfo {author} {\bibfnamefont {Y.}~\bibnamefont {Zhuo}}, \bibinfo {author} {\bibfnamefont {A.}~\bibnamefont {Mansouri~Tehrani}},\ and\ \bibinfo {author} {\bibfnamefont {J.}~\bibnamefont {Brgoch}},\ }\bibfield  {title} {\enquote {\bibinfo {title} {Predicting the {B}and {G}aps of {I}norganic {S}olids by {M}achine {L}earning},}\ }\href {https://doi.org/10.1021/acs.jpclett.8b00124} {\bibfield  {journal} {\bibinfo  {journal} {J. Phys. Chem. Lett.}\ }\textbf {\bibinfo {volume} {9}},\ \bibinfo {pages} {1668–1673} (\bibinfo {year} {2018})}\BibitemShut {NoStop}%
\bibitem [{\citenamefont {Chen}\ \emph {et~al.}(2019)\citenamefont {Chen}, \citenamefont {Ye}, \citenamefont {Zuo}, \citenamefont {Zheng},\ and\ \citenamefont {Ong}}]{Chen2019}%
  \BibitemOpen
  \bibfield  {author} {\bibinfo {author} {\bibfnamefont {C.}~\bibnamefont {Chen}}, \bibinfo {author} {\bibfnamefont {W.}~\bibnamefont {Ye}}, \bibinfo {author} {\bibfnamefont {Y.}~\bibnamefont {Zuo}}, \bibinfo {author} {\bibfnamefont {C.}~\bibnamefont {Zheng}},\ and\ \bibinfo {author} {\bibfnamefont {S.~P.}\ \bibnamefont {Ong}},\ }\bibfield  {title} {\enquote {\bibinfo {title} {Graph {N}etworks as a {U}niversal {M}achine {L}earning {F}ramework for {M}olecules and {C}rystals},}\ }\href {https://doi.org/10.1021/acs.chemmater.9b01294} {\bibfield  {journal} {\bibinfo  {journal} {Chem. Mater.}\ }\textbf {\bibinfo {volume} {31}},\ \bibinfo {pages} {3564–3572} (\bibinfo {year} {2019})}\BibitemShut {NoStop}%
\bibitem [{\citenamefont {Choudhary}\ and\ \citenamefont {DeCost}(2021)}]{Choudhary2021}%
  \BibitemOpen
  \bibfield  {author} {\bibinfo {author} {\bibfnamefont {K.}~\bibnamefont {Choudhary}}\ and\ \bibinfo {author} {\bibfnamefont {B.}~\bibnamefont {DeCost}},\ }\bibfield  {title} {\enquote {\bibinfo {title} {Atomistic {L}ine {G}raph {N}eural {N}etwork for improved materials property predictions},}\ }\href {https://doi.org/10.1038/s41524-021-00650-1} {\bibfield  {journal} {\bibinfo  {journal} {Npj Comput. Mater.}\ }\textbf {\bibinfo {volume} {7}},\ \bibinfo {pages} {185} (\bibinfo {year} {2021})}\BibitemShut {NoStop}%
\bibitem [{\citenamefont {Omee}\ \emph {et~al.}(2022)\citenamefont {Omee}, \citenamefont {Louis}, \citenamefont {Fu}, \citenamefont {Wei}, \citenamefont {Dey}, \citenamefont {Dong}, \citenamefont {Li},\ and\ \citenamefont {Hu}}]{Omee2022}%
  \BibitemOpen
  \bibfield  {author} {\bibinfo {author} {\bibfnamefont {S.~S.}\ \bibnamefont {Omee}}, \bibinfo {author} {\bibfnamefont {S.-Y.}\ \bibnamefont {Louis}}, \bibinfo {author} {\bibfnamefont {N.}~\bibnamefont {Fu}}, \bibinfo {author} {\bibfnamefont {L.}~\bibnamefont {Wei}}, \bibinfo {author} {\bibfnamefont {S.}~\bibnamefont {Dey}}, \bibinfo {author} {\bibfnamefont {R.}~\bibnamefont {Dong}}, \bibinfo {author} {\bibfnamefont {Q.}~\bibnamefont {Li}},\ and\ \bibinfo {author} {\bibfnamefont {J.}~\bibnamefont {Hu}},\ }\bibfield  {title} {\enquote {\bibinfo {title} {Scalable deeper graph neural networks for high-performance materials property prediction},}\ }\href {https://doi.org/10.1016/j.patter.2022.100491} {\bibfield  {journal} {\bibinfo  {journal} {Patterns}\ }\textbf {\bibinfo {volume} {3}},\ \bibinfo {pages} {100491} (\bibinfo {year} {2022})}\BibitemShut {NoStop}%
\bibitem [{\citenamefont {Wang}\ \emph {et~al.}(2022)\citenamefont {Wang}, \citenamefont {Zhang}, \citenamefont {Thé},\ and\ \citenamefont {Yu}}]{Wang2022}%
  \BibitemOpen
  \bibfield  {author} {\bibinfo {author} {\bibfnamefont {T.}~\bibnamefont {Wang}}, \bibinfo {author} {\bibfnamefont {K.}~\bibnamefont {Zhang}}, \bibinfo {author} {\bibfnamefont {J.}~\bibnamefont {Thé}},\ and\ \bibinfo {author} {\bibfnamefont {H.}~\bibnamefont {Yu}},\ }\bibfield  {title} {\enquote {\bibinfo {title} {Accurate prediction of band gap of materials using stacking machine learning model},}\ }\href {https://doi.org/10.1016/j.commatsci.2021.110899} {\bibfield  {journal} {\bibinfo  {journal} {Comput. Mater. Sci.}\ }\textbf {\bibinfo {volume} {201}},\ \bibinfo {pages} {110899} (\bibinfo {year} {2022})}\BibitemShut {NoStop}%
\bibitem [{\citenamefont {Morita}\ \emph {et~al.}(2020)\citenamefont {Morita}, \citenamefont {Davies}, \citenamefont {Butler},\ and\ \citenamefont {Walsh}}]{Morita2020}%
  \BibitemOpen
  \bibfield  {author} {\bibinfo {author} {\bibfnamefont {K.}~\bibnamefont {Morita}}, \bibinfo {author} {\bibfnamefont {D.~W.}\ \bibnamefont {Davies}}, \bibinfo {author} {\bibfnamefont {K.~T.}\ \bibnamefont {Butler}},\ and\ \bibinfo {author} {\bibfnamefont {A.}~\bibnamefont {Walsh}},\ }\bibfield  {title} {\enquote {\bibinfo {title} {Modeling the dielectric constants of crystals using machine learning},}\ }\href {https://doi.org/10.1063/5.0013136} {\bibfield  {journal} {\bibinfo  {journal} {J. Chem. Phys.}\ }\textbf {\bibinfo {volume} {153}} (\bibinfo {year} {2020}),\ 10.1063/5.0013136}\BibitemShut {NoStop}%
\bibitem [{\citenamefont {Takahashi}\ \emph {et~al.}(2020)\citenamefont {Takahashi}, \citenamefont {Kumagai}, \citenamefont {Miyamoto}, \citenamefont {Mochizuki},\ and\ \citenamefont {Oba}}]{Takahashi2020}%
  \BibitemOpen
  \bibfield  {author} {\bibinfo {author} {\bibfnamefont {A.}~\bibnamefont {Takahashi}}, \bibinfo {author} {\bibfnamefont {Y.}~\bibnamefont {Kumagai}}, \bibinfo {author} {\bibfnamefont {J.}~\bibnamefont {Miyamoto}}, \bibinfo {author} {\bibfnamefont {Y.}~\bibnamefont {Mochizuki}},\ and\ \bibinfo {author} {\bibfnamefont {F.}~\bibnamefont {Oba}},\ }\bibfield  {title} {\enquote {\bibinfo {title} {Machine learning models for predicting the dielectric constants of oxides based on high-throughput first-principles calculations},}\ }\href {https://doi.org/10.1103/PhysRevMaterials.4.103801} {\bibfield  {journal} {\bibinfo  {journal} {Phys. Rev. Mater.}\ }\textbf {\bibinfo {volume} {4}},\ \bibinfo {pages} {103801} (\bibinfo {year} {2020})}\BibitemShut {NoStop}%
\bibitem [{\citenamefont {Stanev}\ \emph {et~al.}(2018)\citenamefont {Stanev}, \citenamefont {Oses}, \citenamefont {Kusne}, \citenamefont {Rodriguez}, \citenamefont {Paglione}, \citenamefont {Curtarolo},\ and\ \citenamefont {Takeuchi}}]{Stanev2018}%
  \BibitemOpen
  \bibfield  {author} {\bibinfo {author} {\bibfnamefont {V.}~\bibnamefont {Stanev}}, \bibinfo {author} {\bibfnamefont {C.}~\bibnamefont {Oses}}, \bibinfo {author} {\bibfnamefont {A.~G.}\ \bibnamefont {Kusne}}, \bibinfo {author} {\bibfnamefont {E.}~\bibnamefont {Rodriguez}}, \bibinfo {author} {\bibfnamefont {J.}~\bibnamefont {Paglione}}, \bibinfo {author} {\bibfnamefont {S.}~\bibnamefont {Curtarolo}},\ and\ \bibinfo {author} {\bibfnamefont {I.}~\bibnamefont {Takeuchi}},\ }\bibfield  {title} {\enquote {\bibinfo {title} {Machine learning modeling of superconducting critical temperature},}\ }\href {https://doi.org/10.1038/s41524-018-0085-8} {\bibfield  {journal} {\bibinfo  {journal} {Npj Comput. Mater.}\ }\textbf {\bibinfo {volume} {4}},\ \bibinfo {pages} {29} (\bibinfo {year} {2018})}\BibitemShut {NoStop}%
\bibitem [{\citenamefont {Cerqueira}, \citenamefont {Sanna},\ and\ \citenamefont {Marques}(2023)}]{Cerqueira2023}%
  \BibitemOpen
  \bibfield  {author} {\bibinfo {author} {\bibfnamefont {T.~F.~T.}\ \bibnamefont {Cerqueira}}, \bibinfo {author} {\bibfnamefont {A.}~\bibnamefont {Sanna}},\ and\ \bibinfo {author} {\bibfnamefont {M.~A.~L.}\ \bibnamefont {Marques}},\ }\bibfield  {title} {\enquote {\bibinfo {title} {Sampling the {M}aterials {S}pace for {C}onventional {S}uperconducting {C}ompounds},}\ }\href {https://doi.org/10.1002/adma.202307085} {\bibfield  {journal} {\bibinfo  {journal} {Adv. Mater.}\ }\textbf {\bibinfo {volume} {36}},\ \bibinfo {pages} {2307085} (\bibinfo {year} {2023})}\BibitemShut {NoStop}%
\bibitem [{\citenamefont {Sanna}\ \emph {et~al.}(2024)\citenamefont {Sanna}, \citenamefont {Cerqueira}, \citenamefont {Fang}, \citenamefont {Errea}, \citenamefont {Ludwig},\ and\ \citenamefont {Marques}}]{Sanna2024}%
  \BibitemOpen
  \bibfield  {author} {\bibinfo {author} {\bibfnamefont {A.}~\bibnamefont {Sanna}}, \bibinfo {author} {\bibfnamefont {T.~F.~T.}\ \bibnamefont {Cerqueira}}, \bibinfo {author} {\bibfnamefont {Y.-W.}\ \bibnamefont {Fang}}, \bibinfo {author} {\bibfnamefont {I.}~\bibnamefont {Errea}}, \bibinfo {author} {\bibfnamefont {A.}~\bibnamefont {Ludwig}},\ and\ \bibinfo {author} {\bibfnamefont {M.~A.~L.}\ \bibnamefont {Marques}},\ }\bibfield  {title} {\enquote {\bibinfo {title} {Prediction of ambient pressure conventional superconductivity above {80 K} in hydride compounds},}\ }\href {https://doi.org/10.1038/s41524-024-01214-9} {\bibfield  {journal} {\bibinfo  {journal} {Npj Comput. Mater.}\ }\textbf {\bibinfo {volume} {10}},\ \bibinfo {pages} {44} (\bibinfo {year} {2024})}\BibitemShut {NoStop}%
\bibitem [{\citenamefont {Grunert}, \citenamefont {Großmann},\ and\ \citenamefont {Runge}(2024{\natexlab{a}})}]{Grunert2024b}%
  \BibitemOpen
  \bibfield  {author} {\bibinfo {author} {\bibfnamefont {M.}~\bibnamefont {Grunert}}, \bibinfo {author} {\bibfnamefont {M.}~\bibnamefont {Großmann}},\ and\ \bibinfo {author} {\bibfnamefont {E.}~\bibnamefont {Runge}},\ }\bibfield  {title} {\enquote {\bibinfo {title} {Deep learning of spectra: {P}redicting the dielectric function of semiconductors},}\ }\href {https://doi.org/10.1103/PhysRevMaterials.8.L122201} {\bibfield  {journal} {\bibinfo  {journal} {Phys. Rev. Mater.}\ }\textbf {\bibinfo {volume} {8}},\ \bibinfo {pages} {L122201} (\bibinfo {year} {2024}{\natexlab{a}})}\BibitemShut {NoStop}%
\bibitem [{\citenamefont {Ibrahim}\ and\ \citenamefont {Ataca}(2024)}]{Ibrahim2024}%
  \BibitemOpen
  \bibfield  {author} {\bibinfo {author} {\bibfnamefont {A.}~\bibnamefont {Ibrahim}}\ and\ \bibinfo {author} {\bibfnamefont {C.}~\bibnamefont {Ataca}},\ }\bibfield  {title} {\enquote {\bibinfo {title} {Prediction of {F}requency-{D}ependent {O}ptical {S}pectrum for {S}olid {M}aterials: {A} {M}ultioutput and {M}ultifidelity {M}achine {L}earning {A}pproach},}\ }\href {https://doi.org/10.1021/acsami.4c07328} {\bibfield  {journal} {\bibinfo  {journal} {ACS Appl. Mater. Inter.}\ }\textbf {\bibinfo {volume} {16}},\ \bibinfo {pages} {41145–41156} (\bibinfo {year} {2024})}\BibitemShut {NoStop}%
\bibitem [{\citenamefont {Hung}\ \emph {et~al.}(2024)\citenamefont {Hung}, \citenamefont {Okabe}, \citenamefont {Chotrattanapituk},\ and\ \citenamefont {Li}}]{Hung2024}%
  \BibitemOpen
  \bibfield  {author} {\bibinfo {author} {\bibfnamefont {N.~T.}\ \bibnamefont {Hung}}, \bibinfo {author} {\bibfnamefont {R.}~\bibnamefont {Okabe}}, \bibinfo {author} {\bibfnamefont {A.}~\bibnamefont {Chotrattanapituk}},\ and\ \bibinfo {author} {\bibfnamefont {M.}~\bibnamefont {Li}},\ }\bibfield  {title} {\enquote {\bibinfo {title} {Universal {E}nsemble‐{E}mbedding {G}raph {N}eural {N}etwork for {D}irect {P}rediction of {O}ptical {S}pectra from {C}rystal {S}tructures},}\ }\href {https://doi.org/10.1002/adma.202409175} {\bibfield  {journal} {\bibinfo  {journal} {Adv. Mater.}\ ,\ \bibinfo {pages} {2409175}} (\bibinfo {year} {2024})}\BibitemShut {NoStop}%
\bibitem [{\citenamefont {McInnes}, \citenamefont {Healy},\ and\ \citenamefont {Melville}(2018)}]{UMAP2018}%
  \BibitemOpen
  \bibfield  {author} {\bibinfo {author} {\bibfnamefont {L.}~\bibnamefont {McInnes}}, \bibinfo {author} {\bibfnamefont {J.}~\bibnamefont {Healy}},\ and\ \bibinfo {author} {\bibfnamefont {J.}~\bibnamefont {Melville}},\ }\href {https://doi.org/10.48550/ARXIV.1802.03426} {\enquote {\bibinfo {title} {{UMAP:} {U}niform {M}anifold {A}pproximation and {P}rojection for {D}imension {R}eduction},}\ } (\bibinfo {year} {2018}),\ \Eprint {https://arxiv.org/abs/1802.03426} {arXiv:1802.03426} \BibitemShut {NoStop}%
\bibitem [{\citenamefont {Becht}\ \emph {et~al.}(2018)\citenamefont {Becht}, \citenamefont {McInnes}, \citenamefont {Healy}, \citenamefont {Dutertre}, \citenamefont {Kwok}, \citenamefont {Ng}, \citenamefont {Ginhoux},\ and\ \citenamefont {Newell}}]{Becht2018}%
  \BibitemOpen
  \bibfield  {author} {\bibinfo {author} {\bibfnamefont {E.}~\bibnamefont {Becht}}, \bibinfo {author} {\bibfnamefont {L.}~\bibnamefont {McInnes}}, \bibinfo {author} {\bibfnamefont {J.}~\bibnamefont {Healy}}, \bibinfo {author} {\bibfnamefont {C.-A.}\ \bibnamefont {Dutertre}}, \bibinfo {author} {\bibfnamefont {I.~W.~H.}\ \bibnamefont {Kwok}}, \bibinfo {author} {\bibfnamefont {L.~G.}\ \bibnamefont {Ng}}, \bibinfo {author} {\bibfnamefont {F.}~\bibnamefont {Ginhoux}},\ and\ \bibinfo {author} {\bibfnamefont {E.~W.}\ \bibnamefont {Newell}},\ }\bibfield  {title} {\enquote {\bibinfo {title} {Dimensionality reduction for visualizing single-cell data using {UMAP}},}\ }\href {https://doi.org/10.1038/nbt.4314} {\bibfield  {journal} {\bibinfo  {journal} {Nat. Biotechnol.}\ }\textbf {\bibinfo {volume} {37}},\ \bibinfo {pages} {38–44} (\bibinfo {year} {2018})}\BibitemShut {NoStop}%
\bibitem [{\citenamefont {Karczewski}\ \emph {et~al.}(2020)\citenamefont {Karczewski}, \citenamefont {Francioli}, \citenamefont {Tiao}, \citenamefont {Cummings}, \citenamefont {Alf\"{o}ldi}, \citenamefont {Wang}, \citenamefont {Collins}, \citenamefont {Laricchia}, \citenamefont {Ganna}, \citenamefont {Birnbaum}, \citenamefont {Gauthier}, \citenamefont {Brand}, \citenamefont {Solomonson}, \citenamefont {Watts}, \citenamefont {Rhodes}, \citenamefont {Singer-Berk}, \citenamefont {England}, \citenamefont {Seaby}, \citenamefont {Kosmicki}, \citenamefont {Walters}, \citenamefont {Tashman}, \citenamefont {Farjoun}, \citenamefont {Banks}, \citenamefont {Poterba}, \citenamefont {Wang}, \citenamefont {Seed}, \citenamefont {Whiffin}, \citenamefont {Chong}, \citenamefont {Samocha}, \citenamefont {Pierce-Hoffman}, \citenamefont {Zappala}, \citenamefont {O’Donnell-Luria}, \citenamefont {Minikel}, \citenamefont {Weisburd}, \citenamefont {Lek}, \citenamefont {Ware}, \citenamefont {Vittal}, \citenamefont {Armean},
  \citenamefont {Bergelson}, \citenamefont {Cibulskis}, \citenamefont {Connolly}, \citenamefont {Covarrubias}, \citenamefont {Donnelly}, \citenamefont {Ferriera}, \citenamefont {Gabriel}, \citenamefont {Gentry}, \citenamefont {Gupta}, \citenamefont {Jeandet}, \citenamefont {Kaplan}, \citenamefont {Llanwarne}, \citenamefont {Munshi}, \citenamefont {Novod}, \citenamefont {Petrillo}, \citenamefont {Roazen}, \citenamefont {Ruano-Rubio}, \citenamefont {Saltzman}, \citenamefont {Schleicher}, \citenamefont {Soto}, \citenamefont {Tibbetts}, \citenamefont {Tolonen}, \citenamefont {Wade}, \citenamefont {Talkowski}, \citenamefont {Aguilar~Salinas}, \citenamefont {Ahmad}, \citenamefont {Albert}, \citenamefont {Ardissino}, \citenamefont {Atzmon}, \citenamefont {Barnard}, \citenamefont {Beaugerie}, \citenamefont {Benjamin}, \citenamefont {Boehnke}, \citenamefont {Bonnycastle}, \citenamefont {Bottinger}, \citenamefont {Bowden}, \citenamefont {Bown}, \citenamefont {Chambers}, \citenamefont {Chan}, \citenamefont {Chasman},
  \citenamefont {Cho}, \citenamefont {Chung}, \citenamefont {Cohen}, \citenamefont {Correa}, \citenamefont {Dabelea}, \citenamefont {Daly}, \citenamefont {Darbar}, \citenamefont {Duggirala}, \citenamefont {Dupuis}, \citenamefont {Ellinor}, \citenamefont {Elosua}, \citenamefont {Erdmann}, \citenamefont {Esko}, \citenamefont {F\"{a}rkkil\"{a}}, \citenamefont {Florez}, \citenamefont {Franke}, \citenamefont {Getz}, \citenamefont {Glaser}, \citenamefont {Glatt}, \citenamefont {Goldstein}, \citenamefont {Gonzalez}, \citenamefont {Groop}, \citenamefont {Haiman}, \citenamefont {Hanis}, \citenamefont {Harms}, \citenamefont {Hiltunen}, \citenamefont {Holi}, \citenamefont {Hultman}, \citenamefont {Kallela}, \citenamefont {Kaprio}, \citenamefont {Kathiresan}, \citenamefont {Kim}, \citenamefont {Kim}, \citenamefont {Kirov}, \citenamefont {Kooner}, \citenamefont {Koskinen}, \citenamefont {Krumholz}, \citenamefont {Kugathasan}, \citenamefont {Kwak}, \citenamefont {Laakso}, \citenamefont {Lehtim\"{a}ki}, \citenamefont
  {Loos}, \citenamefont {Lubitz}, \citenamefont {Ma}, \citenamefont {MacArthur}, \citenamefont {Marrugat}, \citenamefont {Mattila}, \citenamefont {McCarroll}, \citenamefont {McCarthy}, \citenamefont {McGovern}, \citenamefont {McPherson}, \citenamefont {Meigs}, \citenamefont {Melander}, \citenamefont {Metspalu}, \citenamefont {Neale}, \citenamefont {Nilsson}, \citenamefont {O’Donovan}, \citenamefont {Ongur}, \citenamefont {Orozco}, \citenamefont {Owen}, \citenamefont {Palmer}, \citenamefont {Palotie}, \citenamefont {Park}, \citenamefont {Pato}, \citenamefont {Pulver}, \citenamefont {Rahman}, \citenamefont {Remes}, \citenamefont {Rioux}, \citenamefont {Ripatti}, \citenamefont {Roden}, \citenamefont {Saleheen}, \citenamefont {Salomaa}, \citenamefont {Samani}, \citenamefont {Scharf}, \citenamefont {Schunkert}, \citenamefont {Shoemaker}, \citenamefont {Sklar}, \citenamefont {Soininen}, \citenamefont {Sokol}, \citenamefont {Spector}, \citenamefont {Sullivan}, \citenamefont {Suvisaari}, \citenamefont {Tai},
  \citenamefont {Teo}, \citenamefont {Tiinamaija}, \citenamefont {Tsuang}, \citenamefont {Turner}, \citenamefont {Tusie-Luna}, \citenamefont {Vartiainen}, \citenamefont {Vawter}, \citenamefont {Ware}, \citenamefont {Watkins}, \citenamefont {Weersma}, \citenamefont {Wessman}, \citenamefont {Wilson}, \citenamefont {Xavier}, \citenamefont {Neale}, \citenamefont {Daly},\ and\ \citenamefont {MacArthur}}]{Karczewski2020}%
  \BibitemOpen
  \bibfield  {author} {\bibinfo {author} {\bibfnamefont {K.~J.}\ \bibnamefont {Karczewski}}, \bibinfo {author} {\bibfnamefont {L.~C.}\ \bibnamefont {Francioli}}, \bibinfo {author} {\bibfnamefont {G.}~\bibnamefont {Tiao}}, \bibinfo {author} {\bibfnamefont {B.~B.}\ \bibnamefont {Cummings}}, \bibinfo {author} {\bibfnamefont {J.}~\bibnamefont {Alf\"{o}ldi}}, \bibinfo {author} {\bibfnamefont {Q.}~\bibnamefont {Wang}}, \bibinfo {author} {\bibfnamefont {R.~L.}\ \bibnamefont {Collins}}, \bibinfo {author} {\bibfnamefont {K.~M.}\ \bibnamefont {Laricchia}}, \bibinfo {author} {\bibfnamefont {A.}~\bibnamefont {Ganna}}, \bibinfo {author} {\bibfnamefont {D.~P.}\ \bibnamefont {Birnbaum}}, \bibinfo {author} {\bibfnamefont {L.~D.}\ \bibnamefont {Gauthier}}, \bibinfo {author} {\bibfnamefont {H.}~\bibnamefont {Brand}}, \bibinfo {author} {\bibfnamefont {M.}~\bibnamefont {Solomonson}}, \bibinfo {author} {\bibfnamefont {N.~A.}\ \bibnamefont {Watts}}, \bibinfo {author} {\bibfnamefont {D.}~\bibnamefont {Rhodes}}, \bibinfo {author}
  {\bibfnamefont {M.}~\bibnamefont {Singer-Berk}}, \bibinfo {author} {\bibfnamefont {E.~M.}\ \bibnamefont {England}}, \bibinfo {author} {\bibfnamefont {E.~G.}\ \bibnamefont {Seaby}}, \bibinfo {author} {\bibfnamefont {J.~A.}\ \bibnamefont {Kosmicki}}, \bibinfo {author} {\bibfnamefont {R.~K.}\ \bibnamefont {Walters}}, \bibinfo {author} {\bibfnamefont {K.}~\bibnamefont {Tashman}}, \bibinfo {author} {\bibfnamefont {Y.}~\bibnamefont {Farjoun}}, \bibinfo {author} {\bibfnamefont {E.}~\bibnamefont {Banks}}, \bibinfo {author} {\bibfnamefont {T.}~\bibnamefont {Poterba}}, \bibinfo {author} {\bibfnamefont {A.}~\bibnamefont {Wang}}, \bibinfo {author} {\bibfnamefont {C.}~\bibnamefont {Seed}}, \bibinfo {author} {\bibfnamefont {N.}~\bibnamefont {Whiffin}}, \bibinfo {author} {\bibfnamefont {J.~X.}\ \bibnamefont {Chong}}, \bibinfo {author} {\bibfnamefont {K.~E.}\ \bibnamefont {Samocha}}, \bibinfo {author} {\bibfnamefont {E.}~\bibnamefont {Pierce-Hoffman}}, \bibinfo {author} {\bibfnamefont {Z.}~\bibnamefont {Zappala}}, \bibinfo
  {author} {\bibfnamefont {A.~H.}\ \bibnamefont {O’Donnell-Luria}}, \bibinfo {author} {\bibfnamefont {E.~V.}\ \bibnamefont {Minikel}}, \bibinfo {author} {\bibfnamefont {B.}~\bibnamefont {Weisburd}}, \bibinfo {author} {\bibfnamefont {M.}~\bibnamefont {Lek}}, \bibinfo {author} {\bibfnamefont {J.~S.}\ \bibnamefont {Ware}}, \bibinfo {author} {\bibfnamefont {C.}~\bibnamefont {Vittal}}, \bibinfo {author} {\bibfnamefont {I.~M.}\ \bibnamefont {Armean}}, \bibinfo {author} {\bibfnamefont {L.}~\bibnamefont {Bergelson}}, \bibinfo {author} {\bibfnamefont {K.}~\bibnamefont {Cibulskis}}, \bibinfo {author} {\bibfnamefont {K.~M.}\ \bibnamefont {Connolly}}, \bibinfo {author} {\bibfnamefont {M.}~\bibnamefont {Covarrubias}}, \bibinfo {author} {\bibfnamefont {S.}~\bibnamefont {Donnelly}}, \bibinfo {author} {\bibfnamefont {S.}~\bibnamefont {Ferriera}}, \bibinfo {author} {\bibfnamefont {S.}~\bibnamefont {Gabriel}}, \bibinfo {author} {\bibfnamefont {J.}~\bibnamefont {Gentry}}, \bibinfo {author} {\bibfnamefont {N.}~\bibnamefont
  {Gupta}}, \bibinfo {author} {\bibfnamefont {T.}~\bibnamefont {Jeandet}}, \bibinfo {author} {\bibfnamefont {D.}~\bibnamefont {Kaplan}}, \bibinfo {author} {\bibfnamefont {C.}~\bibnamefont {Llanwarne}}, \bibinfo {author} {\bibfnamefont {R.}~\bibnamefont {Munshi}}, \bibinfo {author} {\bibfnamefont {S.}~\bibnamefont {Novod}}, \bibinfo {author} {\bibfnamefont {N.}~\bibnamefont {Petrillo}}, \bibinfo {author} {\bibfnamefont {D.}~\bibnamefont {Roazen}}, \bibinfo {author} {\bibfnamefont {V.}~\bibnamefont {Ruano-Rubio}}, \bibinfo {author} {\bibfnamefont {A.}~\bibnamefont {Saltzman}}, \bibinfo {author} {\bibfnamefont {M.}~\bibnamefont {Schleicher}}, \bibinfo {author} {\bibfnamefont {J.}~\bibnamefont {Soto}}, \bibinfo {author} {\bibfnamefont {K.}~\bibnamefont {Tibbetts}}, \bibinfo {author} {\bibfnamefont {C.}~\bibnamefont {Tolonen}}, \bibinfo {author} {\bibfnamefont {G.}~\bibnamefont {Wade}}, \bibinfo {author} {\bibfnamefont {M.~E.}\ \bibnamefont {Talkowski}}, \bibinfo {author} {\bibfnamefont {C.~A.}\ \bibnamefont
  {Aguilar~Salinas}}, \bibinfo {author} {\bibfnamefont {T.}~\bibnamefont {Ahmad}}, \bibinfo {author} {\bibfnamefont {C.~M.}\ \bibnamefont {Albert}}, \bibinfo {author} {\bibfnamefont {D.}~\bibnamefont {Ardissino}}, \bibinfo {author} {\bibfnamefont {G.}~\bibnamefont {Atzmon}}, \bibinfo {author} {\bibfnamefont {J.}~\bibnamefont {Barnard}}, \bibinfo {author} {\bibfnamefont {L.}~\bibnamefont {Beaugerie}}, \bibinfo {author} {\bibfnamefont {E.~J.}\ \bibnamefont {Benjamin}}, \bibinfo {author} {\bibfnamefont {M.}~\bibnamefont {Boehnke}}, \bibinfo {author} {\bibfnamefont {L.~L.}\ \bibnamefont {Bonnycastle}}, \bibinfo {author} {\bibfnamefont {E.~P.}\ \bibnamefont {Bottinger}}, \bibinfo {author} {\bibfnamefont {D.~W.}\ \bibnamefont {Bowden}}, \bibinfo {author} {\bibfnamefont {M.~J.}\ \bibnamefont {Bown}}, \bibinfo {author} {\bibfnamefont {J.~C.}\ \bibnamefont {Chambers}}, \bibinfo {author} {\bibfnamefont {J.~C.}\ \bibnamefont {Chan}}, \bibinfo {author} {\bibfnamefont {D.}~\bibnamefont {Chasman}}, \bibinfo {author}
  {\bibfnamefont {J.}~\bibnamefont {Cho}}, \bibinfo {author} {\bibfnamefont {M.~K.}\ \bibnamefont {Chung}}, \bibinfo {author} {\bibfnamefont {B.}~\bibnamefont {Cohen}}, \bibinfo {author} {\bibfnamefont {A.}~\bibnamefont {Correa}}, \bibinfo {author} {\bibfnamefont {D.}~\bibnamefont {Dabelea}}, \bibinfo {author} {\bibfnamefont {M.~J.}\ \bibnamefont {Daly}}, \bibinfo {author} {\bibfnamefont {D.}~\bibnamefont {Darbar}}, \bibinfo {author} {\bibfnamefont {R.}~\bibnamefont {Duggirala}}, \bibinfo {author} {\bibfnamefont {J.}~\bibnamefont {Dupuis}}, \bibinfo {author} {\bibfnamefont {P.~T.}\ \bibnamefont {Ellinor}}, \bibinfo {author} {\bibfnamefont {R.}~\bibnamefont {Elosua}}, \bibinfo {author} {\bibfnamefont {J.}~\bibnamefont {Erdmann}}, \bibinfo {author} {\bibfnamefont {T.}~\bibnamefont {Esko}}, \bibinfo {author} {\bibfnamefont {M.}~\bibnamefont {F\"{a}rkkil\"{a}}}, \bibinfo {author} {\bibfnamefont {J.}~\bibnamefont {Florez}}, \bibinfo {author} {\bibfnamefont {A.}~\bibnamefont {Franke}}, \bibinfo {author}
  {\bibfnamefont {G.}~\bibnamefont {Getz}}, \bibinfo {author} {\bibfnamefont {B.}~\bibnamefont {Glaser}}, \bibinfo {author} {\bibfnamefont {S.~J.}\ \bibnamefont {Glatt}}, \bibinfo {author} {\bibfnamefont {D.}~\bibnamefont {Goldstein}}, \bibinfo {author} {\bibfnamefont {C.}~\bibnamefont {Gonzalez}}, \bibinfo {author} {\bibfnamefont {L.}~\bibnamefont {Groop}}, \bibinfo {author} {\bibfnamefont {C.}~\bibnamefont {Haiman}}, \bibinfo {author} {\bibfnamefont {C.}~\bibnamefont {Hanis}}, \bibinfo {author} {\bibfnamefont {M.}~\bibnamefont {Harms}}, \bibinfo {author} {\bibfnamefont {M.}~\bibnamefont {Hiltunen}}, \bibinfo {author} {\bibfnamefont {M.~M.}\ \bibnamefont {Holi}}, \bibinfo {author} {\bibfnamefont {C.~M.}\ \bibnamefont {Hultman}}, \bibinfo {author} {\bibfnamefont {M.}~\bibnamefont {Kallela}}, \bibinfo {author} {\bibfnamefont {J.}~\bibnamefont {Kaprio}}, \bibinfo {author} {\bibfnamefont {S.}~\bibnamefont {Kathiresan}}, \bibinfo {author} {\bibfnamefont {B.-J.}\ \bibnamefont {Kim}}, \bibinfo {author}
  {\bibfnamefont {Y.~J.}\ \bibnamefont {Kim}}, \bibinfo {author} {\bibfnamefont {G.}~\bibnamefont {Kirov}}, \bibinfo {author} {\bibfnamefont {J.}~\bibnamefont {Kooner}}, \bibinfo {author} {\bibfnamefont {S.}~\bibnamefont {Koskinen}}, \bibinfo {author} {\bibfnamefont {H.~M.}\ \bibnamefont {Krumholz}}, \bibinfo {author} {\bibfnamefont {S.}~\bibnamefont {Kugathasan}}, \bibinfo {author} {\bibfnamefont {S.~H.}\ \bibnamefont {Kwak}}, \bibinfo {author} {\bibfnamefont {M.}~\bibnamefont {Laakso}}, \bibinfo {author} {\bibfnamefont {T.}~\bibnamefont {Lehtim\"{a}ki}}, \bibinfo {author} {\bibfnamefont {R.~J.~F.}\ \bibnamefont {Loos}}, \bibinfo {author} {\bibfnamefont {S.~A.}\ \bibnamefont {Lubitz}}, \bibinfo {author} {\bibfnamefont {R.~C.~W.}\ \bibnamefont {Ma}}, \bibinfo {author} {\bibfnamefont {D.~G.}\ \bibnamefont {MacArthur}}, \bibinfo {author} {\bibfnamefont {J.}~\bibnamefont {Marrugat}}, \bibinfo {author} {\bibfnamefont {K.~M.}\ \bibnamefont {Mattila}}, \bibinfo {author} {\bibfnamefont {S.}~\bibnamefont
  {McCarroll}}, \bibinfo {author} {\bibfnamefont {M.~I.}\ \bibnamefont {McCarthy}}, \bibinfo {author} {\bibfnamefont {D.}~\bibnamefont {McGovern}}, \bibinfo {author} {\bibfnamefont {R.}~\bibnamefont {McPherson}}, \bibinfo {author} {\bibfnamefont {J.~B.}\ \bibnamefont {Meigs}}, \bibinfo {author} {\bibfnamefont {O.}~\bibnamefont {Melander}}, \bibinfo {author} {\bibfnamefont {A.}~\bibnamefont {Metspalu}}, \bibinfo {author} {\bibfnamefont {B.~M.}\ \bibnamefont {Neale}}, \bibinfo {author} {\bibfnamefont {P.~M.}\ \bibnamefont {Nilsson}}, \bibinfo {author} {\bibfnamefont {M.~C.}\ \bibnamefont {O’Donovan}}, \bibinfo {author} {\bibfnamefont {D.}~\bibnamefont {Ongur}}, \bibinfo {author} {\bibfnamefont {L.}~\bibnamefont {Orozco}}, \bibinfo {author} {\bibfnamefont {M.~J.}\ \bibnamefont {Owen}}, \bibinfo {author} {\bibfnamefont {C.~N.~A.}\ \bibnamefont {Palmer}}, \bibinfo {author} {\bibfnamefont {A.}~\bibnamefont {Palotie}}, \bibinfo {author} {\bibfnamefont {K.~S.}\ \bibnamefont {Park}}, \bibinfo {author} {\bibfnamefont
  {C.}~\bibnamefont {Pato}}, \bibinfo {author} {\bibfnamefont {A.~E.}\ \bibnamefont {Pulver}}, \bibinfo {author} {\bibfnamefont {N.}~\bibnamefont {Rahman}}, \bibinfo {author} {\bibfnamefont {A.~M.}\ \bibnamefont {Remes}}, \bibinfo {author} {\bibfnamefont {J.~D.}\ \bibnamefont {Rioux}}, \bibinfo {author} {\bibfnamefont {S.}~\bibnamefont {Ripatti}}, \bibinfo {author} {\bibfnamefont {D.~M.}\ \bibnamefont {Roden}}, \bibinfo {author} {\bibfnamefont {D.}~\bibnamefont {Saleheen}}, \bibinfo {author} {\bibfnamefont {V.}~\bibnamefont {Salomaa}}, \bibinfo {author} {\bibfnamefont {N.~J.}\ \bibnamefont {Samani}}, \bibinfo {author} {\bibfnamefont {J.}~\bibnamefont {Scharf}}, \bibinfo {author} {\bibfnamefont {H.}~\bibnamefont {Schunkert}}, \bibinfo {author} {\bibfnamefont {M.~B.}\ \bibnamefont {Shoemaker}}, \bibinfo {author} {\bibfnamefont {P.}~\bibnamefont {Sklar}}, \bibinfo {author} {\bibfnamefont {H.}~\bibnamefont {Soininen}}, \bibinfo {author} {\bibfnamefont {H.}~\bibnamefont {Sokol}}, \bibinfo {author} {\bibfnamefont
  {T.}~\bibnamefont {Spector}}, \bibinfo {author} {\bibfnamefont {P.~F.}\ \bibnamefont {Sullivan}}, \bibinfo {author} {\bibfnamefont {J.}~\bibnamefont {Suvisaari}}, \bibinfo {author} {\bibfnamefont {E.~S.}\ \bibnamefont {Tai}}, \bibinfo {author} {\bibfnamefont {Y.~Y.}\ \bibnamefont {Teo}}, \bibinfo {author} {\bibfnamefont {T.}~\bibnamefont {Tiinamaija}}, \bibinfo {author} {\bibfnamefont {M.}~\bibnamefont {Tsuang}}, \bibinfo {author} {\bibfnamefont {D.}~\bibnamefont {Turner}}, \bibinfo {author} {\bibfnamefont {T.}~\bibnamefont {Tusie-Luna}}, \bibinfo {author} {\bibfnamefont {E.}~\bibnamefont {Vartiainen}}, \bibinfo {author} {\bibfnamefont {M.~P.}\ \bibnamefont {Vawter}}, \bibinfo {author} {\bibfnamefont {J.~S.}\ \bibnamefont {Ware}}, \bibinfo {author} {\bibfnamefont {H.}~\bibnamefont {Watkins}}, \bibinfo {author} {\bibfnamefont {R.~K.}\ \bibnamefont {Weersma}}, \bibinfo {author} {\bibfnamefont {M.}~\bibnamefont {Wessman}}, \bibinfo {author} {\bibfnamefont {J.~G.}\ \bibnamefont {Wilson}}, \bibinfo {author}
  {\bibfnamefont {R.~J.}\ \bibnamefont {Xavier}}, \bibinfo {author} {\bibfnamefont {B.~M.}\ \bibnamefont {Neale}}, \bibinfo {author} {\bibfnamefont {M.~J.}\ \bibnamefont {Daly}},\ and\ \bibinfo {author} {\bibfnamefont {D.~G.}\ \bibnamefont {MacArthur}},\ }\bibfield  {title} {\enquote {\bibinfo {title} {The mutational constraint spectrum quantified from variation in 141, 456 humans},}\ }\href {https://doi.org/10.1038/s41586-020-2308-7} {\bibfield  {journal} {\bibinfo  {journal} {Nature}\ }\textbf {\bibinfo {volume} {581}},\ \bibinfo {pages} {434–443} (\bibinfo {year} {2020})}\BibitemShut {NoStop}%
\bibitem [{\citenamefont {Sikkema}\ \emph {et~al.}(2023)\citenamefont {Sikkema}, \citenamefont {Ramírez-Suástegui}, \citenamefont {Strobl}, \citenamefont {Gillett}, \citenamefont {Zappia}, \citenamefont {Madissoon}, \citenamefont {Markov}, \citenamefont {Zaragosi}, \citenamefont {Ji}, \citenamefont {Ansari}, \citenamefont {Arguel}, \citenamefont {Apperloo}, \citenamefont {Banchero}, \citenamefont {Bécavin}, \citenamefont {Berg}, \citenamefont {Chichelnitskiy}, \citenamefont {Chung}, \citenamefont {Collin}, \citenamefont {Gay}, \citenamefont {Gote-Schniering}, \citenamefont {Hooshiar~Kashani}, \citenamefont {Inecik}, \citenamefont {Jain}, \citenamefont {Kapellos}, \citenamefont {Kole}, \citenamefont {Leroy}, \citenamefont {Mayr}, \citenamefont {Oliver}, \citenamefont {von Papen}, \citenamefont {Peter}, \citenamefont {Taylor}, \citenamefont {Walzthoeni}, \citenamefont {Xu}, \citenamefont {Bui}, \citenamefont {De~Donno}, \citenamefont {Dony}, \citenamefont {Faiz}, \citenamefont {Guo}, \citenamefont
  {Gutierrez}, \citenamefont {Heumos}, \citenamefont {Huang}, \citenamefont {Ibarra}, \citenamefont {Jackson}, \citenamefont {Kadur Lakshminarasimha~Murthy}, \citenamefont {Lotfollahi}, \citenamefont {Tabib}, \citenamefont {Talavera-López}, \citenamefont {Travaglini}, \citenamefont {Wilbrey-Clark}, \citenamefont {Worlock}, \citenamefont {Yoshida}, \citenamefont {Chen}, \citenamefont {Hagood}, \citenamefont {Agami}, \citenamefont {Horvath}, \citenamefont {Lundeberg}, \citenamefont {Marquette}, \citenamefont {Pryhuber}, \citenamefont {Samakovlis}, \citenamefont {Sun}, \citenamefont {Ware}, \citenamefont {Zhang}, \citenamefont {van~den Berge}, \citenamefont {Bossé}, \citenamefont {Desai}, \citenamefont {Eickelberg}, \citenamefont {Kaminski}, \citenamefont {Krasnow}, \citenamefont {Lafyatis}, \citenamefont {Nikolic}, \citenamefont {Powell}, \citenamefont {Rajagopal}, \citenamefont {Rojas}, \citenamefont {Rozenblatt-Rosen}, \citenamefont {Seibold}, \citenamefont {Sheppard}, \citenamefont {Shepherd},
  \citenamefont {Sin}, \citenamefont {Timens}, \citenamefont {Tsankov}, \citenamefont {Whitsett}, \citenamefont {Xu}, \citenamefont {Banovich}, \citenamefont {Barbry}, \citenamefont {Duong}, \citenamefont {Falk}, \citenamefont {Meyer}, \citenamefont {Kropski}, \citenamefont {Pe’er}, \citenamefont {Schiller}, \citenamefont {Tata}, \citenamefont {Schultze}, \citenamefont {Teichmann}, \citenamefont {Misharin}, \citenamefont {Nawijn}, \citenamefont {Luecken},\ and\ \citenamefont {Theis}}]{Sikkema2023}%
  \BibitemOpen
  \bibfield  {author} {\bibinfo {author} {\bibfnamefont {L.}~\bibnamefont {Sikkema}}, \bibinfo {author} {\bibfnamefont {C.}~\bibnamefont {Ramírez-Suástegui}}, \bibinfo {author} {\bibfnamefont {D.~C.}\ \bibnamefont {Strobl}}, \bibinfo {author} {\bibfnamefont {T.~E.}\ \bibnamefont {Gillett}}, \bibinfo {author} {\bibfnamefont {L.}~\bibnamefont {Zappia}}, \bibinfo {author} {\bibfnamefont {E.}~\bibnamefont {Madissoon}}, \bibinfo {author} {\bibfnamefont {N.~S.}\ \bibnamefont {Markov}}, \bibinfo {author} {\bibfnamefont {L.-E.}\ \bibnamefont {Zaragosi}}, \bibinfo {author} {\bibfnamefont {Y.}~\bibnamefont {Ji}}, \bibinfo {author} {\bibfnamefont {M.}~\bibnamefont {Ansari}}, \bibinfo {author} {\bibfnamefont {M.-J.}\ \bibnamefont {Arguel}}, \bibinfo {author} {\bibfnamefont {L.}~\bibnamefont {Apperloo}}, \bibinfo {author} {\bibfnamefont {M.}~\bibnamefont {Banchero}}, \bibinfo {author} {\bibfnamefont {C.}~\bibnamefont {Bécavin}}, \bibinfo {author} {\bibfnamefont {M.}~\bibnamefont {Berg}}, \bibinfo {author}
  {\bibfnamefont {E.}~\bibnamefont {Chichelnitskiy}}, \bibinfo {author} {\bibfnamefont {M.-i.}\ \bibnamefont {Chung}}, \bibinfo {author} {\bibfnamefont {A.}~\bibnamefont {Collin}}, \bibinfo {author} {\bibfnamefont {A.~C.~A.}\ \bibnamefont {Gay}}, \bibinfo {author} {\bibfnamefont {J.}~\bibnamefont {Gote-Schniering}}, \bibinfo {author} {\bibfnamefont {B.}~\bibnamefont {Hooshiar~Kashani}}, \bibinfo {author} {\bibfnamefont {K.}~\bibnamefont {Inecik}}, \bibinfo {author} {\bibfnamefont {M.}~\bibnamefont {Jain}}, \bibinfo {author} {\bibfnamefont {T.~S.}\ \bibnamefont {Kapellos}}, \bibinfo {author} {\bibfnamefont {T.~M.}\ \bibnamefont {Kole}}, \bibinfo {author} {\bibfnamefont {S.}~\bibnamefont {Leroy}}, \bibinfo {author} {\bibfnamefont {C.~H.}\ \bibnamefont {Mayr}}, \bibinfo {author} {\bibfnamefont {A.~J.}\ \bibnamefont {Oliver}}, \bibinfo {author} {\bibfnamefont {M.}~\bibnamefont {von Papen}}, \bibinfo {author} {\bibfnamefont {L.}~\bibnamefont {Peter}}, \bibinfo {author} {\bibfnamefont {C.~J.}\ \bibnamefont
  {Taylor}}, \bibinfo {author} {\bibfnamefont {T.}~\bibnamefont {Walzthoeni}}, \bibinfo {author} {\bibfnamefont {C.}~\bibnamefont {Xu}}, \bibinfo {author} {\bibfnamefont {L.~T.}\ \bibnamefont {Bui}}, \bibinfo {author} {\bibfnamefont {C.}~\bibnamefont {De~Donno}}, \bibinfo {author} {\bibfnamefont {L.}~\bibnamefont {Dony}}, \bibinfo {author} {\bibfnamefont {A.}~\bibnamefont {Faiz}}, \bibinfo {author} {\bibfnamefont {M.}~\bibnamefont {Guo}}, \bibinfo {author} {\bibfnamefont {A.~J.}\ \bibnamefont {Gutierrez}}, \bibinfo {author} {\bibfnamefont {L.}~\bibnamefont {Heumos}}, \bibinfo {author} {\bibfnamefont {N.}~\bibnamefont {Huang}}, \bibinfo {author} {\bibfnamefont {I.~L.}\ \bibnamefont {Ibarra}}, \bibinfo {author} {\bibfnamefont {N.~D.}\ \bibnamefont {Jackson}}, \bibinfo {author} {\bibfnamefont {P.}~\bibnamefont {Kadur Lakshminarasimha~Murthy}}, \bibinfo {author} {\bibfnamefont {M.}~\bibnamefont {Lotfollahi}}, \bibinfo {author} {\bibfnamefont {T.}~\bibnamefont {Tabib}}, \bibinfo {author} {\bibfnamefont
  {C.}~\bibnamefont {Talavera-López}}, \bibinfo {author} {\bibfnamefont {K.~J.}\ \bibnamefont {Travaglini}}, \bibinfo {author} {\bibfnamefont {A.}~\bibnamefont {Wilbrey-Clark}}, \bibinfo {author} {\bibfnamefont {K.~B.}\ \bibnamefont {Worlock}}, \bibinfo {author} {\bibfnamefont {M.}~\bibnamefont {Yoshida}}, \bibinfo {author} {\bibfnamefont {Y.}~\bibnamefont {Chen}}, \bibinfo {author} {\bibfnamefont {J.~S.}\ \bibnamefont {Hagood}}, \bibinfo {author} {\bibfnamefont {A.}~\bibnamefont {Agami}}, \bibinfo {author} {\bibfnamefont {P.}~\bibnamefont {Horvath}}, \bibinfo {author} {\bibfnamefont {J.}~\bibnamefont {Lundeberg}}, \bibinfo {author} {\bibfnamefont {C.-H.}\ \bibnamefont {Marquette}}, \bibinfo {author} {\bibfnamefont {G.}~\bibnamefont {Pryhuber}}, \bibinfo {author} {\bibfnamefont {C.}~\bibnamefont {Samakovlis}}, \bibinfo {author} {\bibfnamefont {X.}~\bibnamefont {Sun}}, \bibinfo {author} {\bibfnamefont {L.~B.}\ \bibnamefont {Ware}}, \bibinfo {author} {\bibfnamefont {K.}~\bibnamefont {Zhang}}, \bibinfo {author}
  {\bibfnamefont {M.}~\bibnamefont {van~den Berge}}, \bibinfo {author} {\bibfnamefont {Y.}~\bibnamefont {Bossé}}, \bibinfo {author} {\bibfnamefont {T.~J.}\ \bibnamefont {Desai}}, \bibinfo {author} {\bibfnamefont {O.}~\bibnamefont {Eickelberg}}, \bibinfo {author} {\bibfnamefont {N.}~\bibnamefont {Kaminski}}, \bibinfo {author} {\bibfnamefont {M.~A.}\ \bibnamefont {Krasnow}}, \bibinfo {author} {\bibfnamefont {R.}~\bibnamefont {Lafyatis}}, \bibinfo {author} {\bibfnamefont {M.~Z.}\ \bibnamefont {Nikolic}}, \bibinfo {author} {\bibfnamefont {J.~E.}\ \bibnamefont {Powell}}, \bibinfo {author} {\bibfnamefont {J.}~\bibnamefont {Rajagopal}}, \bibinfo {author} {\bibfnamefont {M.}~\bibnamefont {Rojas}}, \bibinfo {author} {\bibfnamefont {O.}~\bibnamefont {Rozenblatt-Rosen}}, \bibinfo {author} {\bibfnamefont {M.~A.}\ \bibnamefont {Seibold}}, \bibinfo {author} {\bibfnamefont {D.}~\bibnamefont {Sheppard}}, \bibinfo {author} {\bibfnamefont {D.~P.}\ \bibnamefont {Shepherd}}, \bibinfo {author} {\bibfnamefont {D.~D.}\
  \bibnamefont {Sin}}, \bibinfo {author} {\bibfnamefont {W.}~\bibnamefont {Timens}}, \bibinfo {author} {\bibfnamefont {A.~M.}\ \bibnamefont {Tsankov}}, \bibinfo {author} {\bibfnamefont {J.}~\bibnamefont {Whitsett}}, \bibinfo {author} {\bibfnamefont {Y.}~\bibnamefont {Xu}}, \bibinfo {author} {\bibfnamefont {N.~E.}\ \bibnamefont {Banovich}}, \bibinfo {author} {\bibfnamefont {P.}~\bibnamefont {Barbry}}, \bibinfo {author} {\bibfnamefont {T.~E.}\ \bibnamefont {Duong}}, \bibinfo {author} {\bibfnamefont {C.~S.}\ \bibnamefont {Falk}}, \bibinfo {author} {\bibfnamefont {K.~B.}\ \bibnamefont {Meyer}}, \bibinfo {author} {\bibfnamefont {J.~A.}\ \bibnamefont {Kropski}}, \bibinfo {author} {\bibfnamefont {D.}~\bibnamefont {Pe’er}}, \bibinfo {author} {\bibfnamefont {H.~B.}\ \bibnamefont {Schiller}}, \bibinfo {author} {\bibfnamefont {P.~R.}\ \bibnamefont {Tata}}, \bibinfo {author} {\bibfnamefont {J.~L.}\ \bibnamefont {Schultze}}, \bibinfo {author} {\bibfnamefont {S.~A.}\ \bibnamefont {Teichmann}}, \bibinfo {author}
  {\bibfnamefont {A.~V.}\ \bibnamefont {Misharin}}, \bibinfo {author} {\bibfnamefont {M.~C.}\ \bibnamefont {Nawijn}}, \bibinfo {author} {\bibfnamefont {M.~D.}\ \bibnamefont {Luecken}},\ and\ \bibinfo {author} {\bibfnamefont {F.~J.}\ \bibnamefont {Theis}},\ }\bibfield  {title} {\enquote {\bibinfo {title} {An integrated cell atlas of the lung in health and disease},}\ }\href {https://doi.org/10.1038/s41591-023-02327-2} {\bibfield  {journal} {\bibinfo  {journal} {Nat. Med.}\ }\textbf {\bibinfo {volume} {29}},\ \bibinfo {pages} {1563–1577} (\bibinfo {year} {2023})}\BibitemShut {NoStop}%
\bibitem [{\citenamefont {Lin}\ \emph {et~al.}(2023)\citenamefont {Lin}, \citenamefont {Akin}, \citenamefont {Rao}, \citenamefont {Hie}, \citenamefont {Zhu}, \citenamefont {Lu}, \citenamefont {Smetanin}, \citenamefont {Verkuil}, \citenamefont {Kabeli}, \citenamefont {Shmueli}, \citenamefont {dos Santos~Costa}, \citenamefont {Fazel-Zarandi}, \citenamefont {Sercu}, \citenamefont {Candido},\ and\ \citenamefont {Rives}}]{Lin2023}%
  \BibitemOpen
  \bibfield  {author} {\bibinfo {author} {\bibfnamefont {Z.}~\bibnamefont {Lin}}, \bibinfo {author} {\bibfnamefont {H.}~\bibnamefont {Akin}}, \bibinfo {author} {\bibfnamefont {R.}~\bibnamefont {Rao}}, \bibinfo {author} {\bibfnamefont {B.}~\bibnamefont {Hie}}, \bibinfo {author} {\bibfnamefont {Z.}~\bibnamefont {Zhu}}, \bibinfo {author} {\bibfnamefont {W.}~\bibnamefont {Lu}}, \bibinfo {author} {\bibfnamefont {N.}~\bibnamefont {Smetanin}}, \bibinfo {author} {\bibfnamefont {R.}~\bibnamefont {Verkuil}}, \bibinfo {author} {\bibfnamefont {O.}~\bibnamefont {Kabeli}}, \bibinfo {author} {\bibfnamefont {Y.}~\bibnamefont {Shmueli}}, \bibinfo {author} {\bibfnamefont {A.}~\bibnamefont {dos Santos~Costa}}, \bibinfo {author} {\bibfnamefont {M.}~\bibnamefont {Fazel-Zarandi}}, \bibinfo {author} {\bibfnamefont {T.}~\bibnamefont {Sercu}}, \bibinfo {author} {\bibfnamefont {S.}~\bibnamefont {Candido}},\ and\ \bibinfo {author} {\bibfnamefont {A.}~\bibnamefont {Rives}},\ }\bibfield  {title} {\enquote {\bibinfo {title}
  {Evolutionary-scale prediction of atomic-level protein structure with a language model},}\ }\href {https://doi.org/10.1126/science.ade2574} {\bibfield  {journal} {\bibinfo  {journal} {Science}\ }\textbf {\bibinfo {volume} {379}},\ \bibinfo {pages} {1123–1130} (\bibinfo {year} {2023})}\BibitemShut {NoStop}%
\bibitem [{\citenamefont {{European Commission}}\ \emph {et~al.}(2023)\citenamefont {{European Commission}}, \citenamefont {{Joint Research Centre}}, \citenamefont {Carrara}, \citenamefont {Bobba}, \citenamefont {Blagoeva}, \citenamefont {Alves~Dias}, \citenamefont {Cavalli}, \citenamefont {Georgitzikis}, \citenamefont {Grohol}, \citenamefont {Itul}, \citenamefont {Kuzov}, \citenamefont {Latunussa}, \citenamefont {Lyons}, \citenamefont {Malano}, \citenamefont {Maury}, \citenamefont {Prior~Arce}, \citenamefont {Somers}, \citenamefont {Telsnig}, \citenamefont {Veeh}, \citenamefont {Wittmer}, \citenamefont {Black}, \citenamefont {Pennington},\ and\ \citenamefont {Christou}}]{EuCrit}%
  \BibitemOpen
  \bibfield  {author} {\bibinfo {author} {\bibnamefont {{European Commission}}}, \bibinfo {author} {\bibnamefont {{Joint Research Centre}}}, \bibinfo {author} {\bibfnamefont {S.}~\bibnamefont {Carrara}}, \bibinfo {author} {\bibfnamefont {S.}~\bibnamefont {Bobba}}, \bibinfo {author} {\bibfnamefont {D.}~\bibnamefont {Blagoeva}}, \bibinfo {author} {\bibfnamefont {P.}~\bibnamefont {Alves~Dias}}, \bibinfo {author} {\bibfnamefont {A.}~\bibnamefont {Cavalli}}, \bibinfo {author} {\bibfnamefont {K.}~\bibnamefont {Georgitzikis}}, \bibinfo {author} {\bibfnamefont {M.}~\bibnamefont {Grohol}}, \bibinfo {author} {\bibfnamefont {A.}~\bibnamefont {Itul}}, \bibinfo {author} {\bibfnamefont {T.}~\bibnamefont {Kuzov}}, \bibinfo {author} {\bibfnamefont {C.}~\bibnamefont {Latunussa}}, \bibinfo {author} {\bibfnamefont {L.}~\bibnamefont {Lyons}}, \bibinfo {author} {\bibfnamefont {G.}~\bibnamefont {Malano}}, \bibinfo {author} {\bibfnamefont {T.}~\bibnamefont {Maury}}, \bibinfo {author} {\bibfnamefont {A.}~\bibnamefont {Prior~Arce}},
  \bibinfo {author} {\bibfnamefont {J.}~\bibnamefont {Somers}}, \bibinfo {author} {\bibfnamefont {T.}~\bibnamefont {Telsnig}}, \bibinfo {author} {\bibfnamefont {C.}~\bibnamefont {Veeh}}, \bibinfo {author} {\bibfnamefont {D.}~\bibnamefont {Wittmer}}, \bibinfo {author} {\bibfnamefont {C.}~\bibnamefont {Black}}, \bibinfo {author} {\bibfnamefont {D.}~\bibnamefont {Pennington}},\ and\ \bibinfo {author} {\bibfnamefont {M.}~\bibnamefont {Christou}},\ }\href {https://doi.org/doi/10.2760/386650} {\emph {\bibinfo {title} {Supply chain analysis and material demand forecast in strategic technologies and sectors in the {EU} – {A} foresight study}}}\ (\bibinfo  {publisher} {Publications Office of the European Union},\ \bibinfo {year} {2023})\BibitemShut {NoStop}%
\bibitem [{\citenamefont {{European Chemical Society}}()}]{euchemsElementScarcity}%
  \BibitemOpen
  \bibfield  {author} {\bibinfo {author} {\bibnamefont {{European Chemical Society}}},\ }\href@noop {} {\enquote {\bibinfo {title} {{E}lement {S}carcity - {E}u{C}hem{S} {P}eriodic {T}able - {E}u{C}hem{S} --- euchems.eu},}\ }\bibinfo {howpublished} {\url{https://www.euchems.eu/euchems-periodic-table/}},\ \bibinfo {note} {[Accessed 20-11-2024]}\BibitemShut {NoStop}%
\bibitem [{\citenamefont {Nuss}\ and\ \citenamefont {Eckelman}(2014)}]{Nuss2014}%
  \BibitemOpen
  \bibfield  {author} {\bibinfo {author} {\bibfnamefont {P.}~\bibnamefont {Nuss}}\ and\ \bibinfo {author} {\bibfnamefont {M.}~\bibnamefont {Eckelman}},\ }\bibfield  {title} {\enquote {\bibinfo {title} {Life cycle assessment of metals: A scientific synthesis},}\ }\href {https://doi.org/10.1371/journal.pone.0101298} {\bibfield  {journal} {\bibinfo  {journal} {PLOS ONE}\ }\textbf {\bibinfo {volume} {9}},\ \bibinfo {pages} {e101298} (\bibinfo {year} {2014})}\BibitemShut {NoStop}%
\bibitem [{\citenamefont {Schrijvers}\ \emph {et~al.}(2020)\citenamefont {Schrijvers}, \citenamefont {Hool}, \citenamefont {Blengini}, \citenamefont {Chen}, \citenamefont {Dewulf}, \citenamefont {Eggert}, \citenamefont {{van Ellen}}, \citenamefont {Gauss}, \citenamefont {Goddin}, \citenamefont {Habib}, \citenamefont {Hagelüken}, \citenamefont {Hirohata}, \citenamefont {Hofmann-Amtenbrink}, \citenamefont {Kosmol}, \citenamefont {{Le Gleuher}}, \citenamefont {Grohol}, \citenamefont {Ku}, \citenamefont {Lee}, \citenamefont {Liu}, \citenamefont {Nansai}, \citenamefont {Nuss}, \citenamefont {Peck}, \citenamefont {Reller}, \citenamefont {Sonnemann}, \citenamefont {Tercero}, \citenamefont {Thorenz},\ and\ \citenamefont {Wäger}}]{Schrijvers2020}%
  \BibitemOpen
  \bibfield  {author} {\bibinfo {author} {\bibfnamefont {D.}~\bibnamefont {Schrijvers}}, \bibinfo {author} {\bibfnamefont {A.}~\bibnamefont {Hool}}, \bibinfo {author} {\bibfnamefont {G.~A.}\ \bibnamefont {Blengini}}, \bibinfo {author} {\bibfnamefont {W.-Q.}\ \bibnamefont {Chen}}, \bibinfo {author} {\bibfnamefont {J.}~\bibnamefont {Dewulf}}, \bibinfo {author} {\bibfnamefont {R.}~\bibnamefont {Eggert}}, \bibinfo {author} {\bibfnamefont {L.}~\bibnamefont {{van Ellen}}}, \bibinfo {author} {\bibfnamefont {R.}~\bibnamefont {Gauss}}, \bibinfo {author} {\bibfnamefont {J.}~\bibnamefont {Goddin}}, \bibinfo {author} {\bibfnamefont {K.}~\bibnamefont {Habib}}, \bibinfo {author} {\bibfnamefont {C.}~\bibnamefont {Hagelüken}}, \bibinfo {author} {\bibfnamefont {A.}~\bibnamefont {Hirohata}}, \bibinfo {author} {\bibfnamefont {M.}~\bibnamefont {Hofmann-Amtenbrink}}, \bibinfo {author} {\bibfnamefont {J.}~\bibnamefont {Kosmol}}, \bibinfo {author} {\bibfnamefont {M.}~\bibnamefont {{Le Gleuher}}}, \bibinfo {author} {\bibfnamefont
  {M.}~\bibnamefont {Grohol}}, \bibinfo {author} {\bibfnamefont {A.}~\bibnamefont {Ku}}, \bibinfo {author} {\bibfnamefont {M.-H.}\ \bibnamefont {Lee}}, \bibinfo {author} {\bibfnamefont {G.}~\bibnamefont {Liu}}, \bibinfo {author} {\bibfnamefont {K.}~\bibnamefont {Nansai}}, \bibinfo {author} {\bibfnamefont {P.}~\bibnamefont {Nuss}}, \bibinfo {author} {\bibfnamefont {D.}~\bibnamefont {Peck}}, \bibinfo {author} {\bibfnamefont {A.}~\bibnamefont {Reller}}, \bibinfo {author} {\bibfnamefont {G.}~\bibnamefont {Sonnemann}}, \bibinfo {author} {\bibfnamefont {L.}~\bibnamefont {Tercero}}, \bibinfo {author} {\bibfnamefont {A.}~\bibnamefont {Thorenz}},\ and\ \bibinfo {author} {\bibfnamefont {P.~A.}\ \bibnamefont {Wäger}},\ }\bibfield  {title} {\enquote {\bibinfo {title} {A review of methods and data to determine raw material criticality},}\ }\href {https://doi.org/https://doi.org/10.1016/j.resconrec.2019.104617} {\bibfield  {journal} {\bibinfo  {journal} {Resour. Conserv. Recycl.}\ }\textbf {\bibinfo {volume} {155}},\
  \bibinfo {pages} {104617} (\bibinfo {year} {2020})}\BibitemShut {NoStop}%
\bibitem [{\citenamefont {Pashley}(1989)}]{Pashley1989}%
  \BibitemOpen
  \bibfield  {author} {\bibinfo {author} {\bibfnamefont {M.~D.}\ \bibnamefont {Pashley}},\ }\bibfield  {title} {\enquote {\bibinfo {title} {Electron counting model and its application to island structures on molecular-beam epitaxy grown {GaAs}(001) and {ZnSe}(001)},}\ }\href {https://doi.org/10.1103/PhysRevB.40.10481} {\bibfield  {journal} {\bibinfo  {journal} {Phys. Rev. B}\ }\textbf {\bibinfo {volume} {40}},\ \bibinfo {pages} {10481--10487} (\bibinfo {year} {1989})}\BibitemShut {NoStop}%
\bibitem [{\citenamefont {Grunert}, \citenamefont {Großmann},\ and\ \citenamefont {Runge}(2024{\natexlab{b}})}]{Grunert2024a}%
  \BibitemOpen
  \bibfield  {author} {\bibinfo {author} {\bibfnamefont {M.}~\bibnamefont {Grunert}}, \bibinfo {author} {\bibfnamefont {M.}~\bibnamefont {Großmann}},\ and\ \bibinfo {author} {\bibfnamefont {E.}~\bibnamefont {Runge}},\ }\bibfield  {title} {\enquote {\bibinfo {title} {Predicting exciton binding energies from ground-state properties},}\ }\href {https://doi.org/10.1103/PhysRevB.110.075204} {\bibfield  {journal} {\bibinfo  {journal} {Phys. Rev. B}\ }\textbf {\bibinfo {volume} {110}},\ \bibinfo {pages} {075204} (\bibinfo {year} {2024}{\natexlab{b}})}\BibitemShut {NoStop}%
\bibitem [{\citenamefont {Gunn}(2013)}]{Gunn2013-uc}%
  \BibitemOpen
  \bibinfo {editor} {\bibfnamefont {G.}~\bibnamefont {Gunn}},\ ed.,\ \href@noop {} {\emph {\bibinfo {title} {Critical metals handbook}}}\ (\bibinfo  {publisher} {Wiley-Blackwell},\ \bibinfo {year} {2013})\BibitemShut {NoStop}%
\bibitem [{\citenamefont {Ratnaike}(2003)}]{Ratnaike2003}%
  \BibitemOpen
  \bibfield  {author} {\bibinfo {author} {\bibfnamefont {R.~N.}\ \bibnamefont {Ratnaike}},\ }\bibfield  {title} {\enquote {\bibinfo {title} {Acute and chronic arsenic toxicity},}\ }\href {https://doi.org/10.1136/pmj.79.933.391} {\bibfield  {journal} {\bibinfo  {journal} {Postgrad. Med. J.}\ }\textbf {\bibinfo {volume} {79}},\ \bibinfo {pages} {391–396} (\bibinfo {year} {2003})}\BibitemShut {NoStop}%
\bibitem [{\citenamefont {Glavič}\ and\ \citenamefont {Lukman}(2007)}]{Glavi2007}%
  \BibitemOpen
  \bibfield  {author} {\bibinfo {author} {\bibfnamefont {P.}~\bibnamefont {Glavič}}\ and\ \bibinfo {author} {\bibfnamefont {R.}~\bibnamefont {Lukman}},\ }\bibfield  {title} {\enquote {\bibinfo {title} {Review of sustainability terms and their definitions},}\ }\href {https://doi.org/10.1016/j.jclepro.2006.12.006} {\bibfield  {journal} {\bibinfo  {journal} {J. Clean. Prod.}\ }\textbf {\bibinfo {volume} {15}},\ \bibinfo {pages} {1875–1885} (\bibinfo {year} {2007})}\BibitemShut {NoStop}%
\bibitem [{\citenamefont {Schmidt}\ \emph {et~al.}(2024)\citenamefont {Schmidt}, \citenamefont {Cerqueira}, \citenamefont {Romero}, \citenamefont {Loew}, \citenamefont {J\"{a}ger}, \citenamefont {Wang}, \citenamefont {Botti},\ and\ \citenamefont {Marques}}]{Schmidt2024}%
  \BibitemOpen
  \bibfield  {author} {\bibinfo {author} {\bibfnamefont {J.}~\bibnamefont {Schmidt}}, \bibinfo {author} {\bibfnamefont {T.~F.}\ \bibnamefont {Cerqueira}}, \bibinfo {author} {\bibfnamefont {A.~H.}\ \bibnamefont {Romero}}, \bibinfo {author} {\bibfnamefont {A.}~\bibnamefont {Loew}}, \bibinfo {author} {\bibfnamefont {F.}~\bibnamefont {J\"{a}ger}}, \bibinfo {author} {\bibfnamefont {H.-C.}\ \bibnamefont {Wang}}, \bibinfo {author} {\bibfnamefont {S.}~\bibnamefont {Botti}},\ and\ \bibinfo {author} {\bibfnamefont {M.~A.}\ \bibnamefont {Marques}},\ }\bibfield  {title} {\enquote {\bibinfo {title} {Improving machine-learning models in materials science through large datasets},}\ }\href {https://doi.org/10.1016/j.mtphys.2024.101560} {\bibfield  {journal} {\bibinfo  {journal} {Mater. Today Phys.}\ }\textbf {\bibinfo {volume} {48}},\ \bibinfo {pages} {101560} (\bibinfo {year} {2024})}\BibitemShut {NoStop}%
\bibitem [{\citenamefont {Hannappel}\ \emph {et~al.}(2024)\citenamefont {Hannappel}, \citenamefont {Shekarabi}, \citenamefont {Jaegermann}, \citenamefont {Runge}, \citenamefont {Hofmann}, \citenamefont {van~de Krol}, \citenamefont {May}, \citenamefont {Paszuk}, \citenamefont {Hess}, \citenamefont {Bergmann}, \citenamefont {Bund}, \citenamefont {Cierpka}, \citenamefont {Dreßler}, \citenamefont {Dionigi}, \citenamefont {Friedrich}, \citenamefont {Favaro}, \citenamefont {Krischok}, \citenamefont {Kurniawan}, \citenamefont {L\"{u}dge}, \citenamefont {Lei}, \citenamefont {Roldán~Cuenya}, \citenamefont {Schaaf}, \citenamefont {Schmidt‐Grund}, \citenamefont {Schmidt}, \citenamefont {Strasser}, \citenamefont {Unger}, \citenamefont {Vasquez~Montoya}, \citenamefont {Wang},\ and\ \citenamefont {Zhang}}]{Hannappel2024}%
  \BibitemOpen
  \bibfield  {author} {\bibinfo {author} {\bibfnamefont {T.}~\bibnamefont {Hannappel}}, \bibinfo {author} {\bibfnamefont {S.}~\bibnamefont {Shekarabi}}, \bibinfo {author} {\bibfnamefont {W.}~\bibnamefont {Jaegermann}}, \bibinfo {author} {\bibfnamefont {E.}~\bibnamefont {Runge}}, \bibinfo {author} {\bibfnamefont {J.~P.}\ \bibnamefont {Hofmann}}, \bibinfo {author} {\bibfnamefont {R.}~\bibnamefont {van~de Krol}}, \bibinfo {author} {\bibfnamefont {M.~M.}\ \bibnamefont {May}}, \bibinfo {author} {\bibfnamefont {A.}~\bibnamefont {Paszuk}}, \bibinfo {author} {\bibfnamefont {F.}~\bibnamefont {Hess}}, \bibinfo {author} {\bibfnamefont {A.}~\bibnamefont {Bergmann}}, \bibinfo {author} {\bibfnamefont {A.}~\bibnamefont {Bund}}, \bibinfo {author} {\bibfnamefont {C.}~\bibnamefont {Cierpka}}, \bibinfo {author} {\bibfnamefont {C.}~\bibnamefont {Dreßler}}, \bibinfo {author} {\bibfnamefont {F.}~\bibnamefont {Dionigi}}, \bibinfo {author} {\bibfnamefont {D.}~\bibnamefont {Friedrich}}, \bibinfo {author} {\bibfnamefont
  {M.}~\bibnamefont {Favaro}}, \bibinfo {author} {\bibfnamefont {S.}~\bibnamefont {Krischok}}, \bibinfo {author} {\bibfnamefont {M.}~\bibnamefont {Kurniawan}}, \bibinfo {author} {\bibfnamefont {K.}~\bibnamefont {L\"{u}dge}}, \bibinfo {author} {\bibfnamefont {Y.}~\bibnamefont {Lei}}, \bibinfo {author} {\bibfnamefont {B.}~\bibnamefont {Roldán~Cuenya}}, \bibinfo {author} {\bibfnamefont {P.}~\bibnamefont {Schaaf}}, \bibinfo {author} {\bibfnamefont {R.}~\bibnamefont {Schmidt‐Grund}}, \bibinfo {author} {\bibfnamefont {W.~G.}\ \bibnamefont {Schmidt}}, \bibinfo {author} {\bibfnamefont {P.}~\bibnamefont {Strasser}}, \bibinfo {author} {\bibfnamefont {E.}~\bibnamefont {Unger}}, \bibinfo {author} {\bibfnamefont {M.~F.}\ \bibnamefont {Vasquez~Montoya}}, \bibinfo {author} {\bibfnamefont {D.}~\bibnamefont {Wang}},\ and\ \bibinfo {author} {\bibfnamefont {H.}~\bibnamefont {Zhang}},\ }\bibfield  {title} {\enquote {\bibinfo {title} {Integration of {M}ultijunction {A}bsorbers and {C}atalysts for {E}fficient {S}olar‐{D}riven
  {A}rtificial {L}eaf {S}tructures: {A} {P}hysical and {M}aterials {S}cience {P}erspective},}\ }\href {https://doi.org/10.1002/solr.202301047} {\bibfield  {journal} {\bibinfo  {journal} {Sol. RRL}\ }\textbf {\bibinfo {volume} {8}},\ \bibinfo {pages} {2301047} (\bibinfo {year} {2024})}\BibitemShut {NoStop}%
\bibitem [{\citenamefont {Jain}\ \emph {et~al.}(2013)\citenamefont {Jain}, \citenamefont {Ong}, \citenamefont {Hautier}, \citenamefont {Chen}, \citenamefont {Richards}, \citenamefont {Dacek}, \citenamefont {Cholia}, \citenamefont {Gunter}, \citenamefont {Skinner}, \citenamefont {Ceder},\ and\ \citenamefont {Persson}}]{Jain2013}%
  \BibitemOpen
  \bibfield  {author} {\bibinfo {author} {\bibfnamefont {A.}~\bibnamefont {Jain}}, \bibinfo {author} {\bibfnamefont {S.~P.}\ \bibnamefont {Ong}}, \bibinfo {author} {\bibfnamefont {G.}~\bibnamefont {Hautier}}, \bibinfo {author} {\bibfnamefont {W.}~\bibnamefont {Chen}}, \bibinfo {author} {\bibfnamefont {W.~D.}\ \bibnamefont {Richards}}, \bibinfo {author} {\bibfnamefont {S.}~\bibnamefont {Dacek}}, \bibinfo {author} {\bibfnamefont {S.}~\bibnamefont {Cholia}}, \bibinfo {author} {\bibfnamefont {D.}~\bibnamefont {Gunter}}, \bibinfo {author} {\bibfnamefont {D.}~\bibnamefont {Skinner}}, \bibinfo {author} {\bibfnamefont {G.}~\bibnamefont {Ceder}},\ and\ \bibinfo {author} {\bibfnamefont {K.~A.}\ \bibnamefont {Persson}},\ }\bibfield  {title} {\enquote {\bibinfo {title} {Commentary: {T}he {M}aterials {P}roject: {A} materials genome approach to accelerating materials innovation},}\ }\href {https://doi.org/10.1063/1.4812323} {\bibfield  {journal} {\bibinfo  {journal} {APL Mater.}\ }\textbf {\bibinfo {volume} {1}},\ \bibinfo
  {pages} {011002} (\bibinfo {year} {2013})}\BibitemShut {NoStop}%
\bibitem [{\citenamefont {Barroso-Luque}\ \emph {et~al.}(2024)\citenamefont {Barroso-Luque}, \citenamefont {Shuaibi}, \citenamefont {Fu}, \citenamefont {Wood}, \citenamefont {Dzamba}, \citenamefont {Gao}, \citenamefont {Rizvi}, \citenamefont {Zitnick},\ and\ \citenamefont {Ulissi}}]{Barroso-Luque2024}%
  \BibitemOpen
  \bibfield  {author} {\bibinfo {author} {\bibfnamefont {L.}~\bibnamefont {Barroso-Luque}}, \bibinfo {author} {\bibfnamefont {M.}~\bibnamefont {Shuaibi}}, \bibinfo {author} {\bibfnamefont {X.}~\bibnamefont {Fu}}, \bibinfo {author} {\bibfnamefont {B.~M.}\ \bibnamefont {Wood}}, \bibinfo {author} {\bibfnamefont {M.}~\bibnamefont {Dzamba}}, \bibinfo {author} {\bibfnamefont {M.}~\bibnamefont {Gao}}, \bibinfo {author} {\bibfnamefont {A.}~\bibnamefont {Rizvi}}, \bibinfo {author} {\bibfnamefont {C.~L.}\ \bibnamefont {Zitnick}},\ and\ \bibinfo {author} {\bibfnamefont {Z.~W.}\ \bibnamefont {Ulissi}},\ }\bibfield  {title} {\enquote {\bibinfo {title} {{O}pen {M}aterials 2024 (omat24) inorganic materials dataset and models},}\ }\href@noop {} {\bibfield  {journal} {\bibinfo  {journal} {arXiv:2410.12771}\ } (\bibinfo {year} {2024})}\BibitemShut {NoStop}%
\end{thebibliography}
\end{document}